\documentclass[lettersize,journal]{IEEEtran}
\usepackage{amsmath,amsfonts}
\usepackage{algorithm}
\usepackage{algpseudocode}
\usepackage{array}
\usepackage{bm}
\usepackage[caption=false,font=normalsize,labelfont=sf,textfont=sf]{subfig}
\usepackage{textcomp}
\usepackage{stfloats}
\usepackage{url}
\usepackage{soul}
\usepackage{verbatim}
\usepackage{graphicx}
\usepackage{cite}
\usepackage{xcolor}
\usepackage{booktabs}
\usepackage{makecell}
\usepackage[flushleft]{threeparttable}
\usepackage[colorlinks,
    linkcolor=red,citecolor=citecolor,urlcolor=blue]{hyperref}       
\newcolumntype{C}[1]{>{\centering\arraybackslash}p{#1}}

\usepackage{tikz,xcolor,hyperref}
\definecolor{lime}{HTML}{A6CE39}
\DeclareRobustCommand{\orcidicon}{
\begin{tikzpicture}
\draw[lime, fill=lime] (0,0)
circle[radius=0.16]
node[white]{{\fontfamily{qag}\selectfont \tiny \.{I}D}};
\end{tikzpicture}
\hspace{-2mm}
}
\foreach \x in {A, ..., Z}{%
\expandafter\xdef\csname orcid\x\endcsname{\noexpand\href{https://orcid.org/\csname orcidauthor\x\endcsname}{\noexpand\orcidicon}}
}

\begin{document}
\definecolor{mypurple}{RGB}{111,0,255}
\definecolor{citecolor}{RGB}{34,139,34}

\newcommand{\JG}[1]{\textcolor{blue}{[JG: #1]}}
\newcommand{\HZ}[1]{\textcolor{mypurple}{[HZ: #1]}}
\newcommand{\ZX}[1]{\textcolor{orange}{[ZX: #1]}}
\newcommand{\CF}[1]{\textcolor{green}{[CF: #1]}}
\newcommand{\SN}[1]{\textcolor{pink}{[SN: #1]}}

\title{Photonic-Electronic Integrated Circuits for High-Performance Computing and AI Accelerators}
\author{
Shupeng Ning\IEEEauthorrefmark{1}\orcidA{},
Hanqing Zhu\IEEEauthorrefmark{1},
Chenghao Feng,
Jiaqi Gu,
Zhixing Jiang,
Zhoufeng Ying,\\ \vspace{1pt}
Jason Midkiff,
Sourabh Jain,
May H. Hlaing,
David Z. Pan,~\IEEEmembership{Fellow,~IEEE},
and
Ray T. Chen\IEEEauthorrefmark{2},~\IEEEmembership{Fellow,~IEEE}
\thanks{S. Ning, H. Zhu, C. Feng, Z. Jiang, Z. Ying, J. Midkiff, S. Jain, M. H. Hlaing, D. Z. Pan and R. T. Chen are with the Department of Electrical and Computer Engineering, The University of Texas at Austin, TX 78712 USA.}
\thanks{J. Gu is with the Department of Electrical and Computer Engineering, The University of Texas at Austin, TX 78712 USA, and also with School of Electrical, Computer and Energy Engineering, Arizona State University, Tempe, AZ 85281 USA.}
\thanks{\IEEEauthorrefmark{1} Both authors are recognized as co-first authors.}
\thanks{\IEEEauthorrefmark{2} Corresponding author: Ray T. Chen (email: chenrt@austin.utexas.edu)}
}
\IEEEaftertitletext{\vspace{-2\baselineskip}\noindent\centerline{\large{(\textit{IEEE Preprint})}}\par\vspace{\baselineskip}}
\vspace{5pt}

\markboth { }%
{Shell \MakeLowercase{\textit{et al.}}: A Sample Article Using IEEEtran.cls for IEEE Journals}


\maketitle

\begin{abstract}
In recent decades, the demand for computational power has surged, particularly with the rapid expansion of artificial intelligence (AI). As we navigate the post-Moore's law era, the limitations of traditional electrical digital computing, including process bottlenecks and power consumption issues, are propelling the search for alternative computing paradigms. Among various emerging technologies, integrated photonics stands out as a promising solution for next-generation high-performance computing, thanks to the inherent advantages of light, such as low latency, high bandwidth, and unique multiplexing techniques. 
Furthermore, the progress in photonic integrated circuits (PICs), which are equipped with abundant photoelectronic components, positions photonic-electronic integrated circuits as a viable solution for high-performance computing and hardware AI accelerators. In this review, we survey recent advancements in both PIC-based digital and analog computing for AI, exploring the principal benefits and obstacles of implementation. Additionally, we propose a comprehensive analysis of photonic AI from the perspectives of hardware implementation, accelerator architecture, and software-hardware co-design. In the end, acknowledging the existing challenges, we underscore potential strategies for overcoming these issues and offer insights into the future drivers for optical computing.
\end{abstract}


\section{Introduction}\label{intro}

\IEEEPARstart{A}{s} the semiconductor industry advances to process nodes below 3 nanometers, it increasingly encounters inherent physical limitations of both devices and materials~\cite{fang2019towards, dennard1974design}. A primary concern is the surge in power consumption as clock frequencies reach gigahertz levels, leading to overwhelming heat generation~\cite{dennard1974design,waldrop2016more}. Furthermore, at these diminutive scales, quantum uncertainties begin to dominate electron behavior, resulting in increased transistor errors and reduced reliability. Additionally, AI has made remarkable strides in recent years, exerting a growing influence on various aspects of our lives, such as image recognition \cite{lecun2015deep,simonyan2014very,redmon2016you,he2016deep}, natural language processing \cite{devlin2018bert,vaswani2017attention}, autonomous driving \cite{bojarski2016end}, and medical diagnosis \cite{hannun2019cardiologist,cao2023large}, which have further increased societal demand for computational power.
One notable example is the emergence of large language models (LLMs) such as GPT (Generative Pre-trained Transformer). These models exhibit human-level intelligence and have revolutionized a wide range of applications, from sophisticated chatbots to advanced text analysis tools.
However, the advancements of deep neural networks (DNNs) are driven by rapidly increasing model sizes and data volumes, which necessitate significantly expanding computational demands. For instance, the GPT-3 model developed by OpenAI, which contains around 175 billion parameters, requires 14.8 days for training using a cluster of around 10,000 NVIDIA V100 GPUs, with an estimated energy consumption of 1287 MWh \cite{brown2020language,patterson2021carbon,epochai2024notable}.
Hence, in the post-Moore's Law era, traditional electronic computing architectures, designed to execute sequential, digital programs, are inadequate to meet the surging demand for high-performance computing and AI tasks.
There is a pressing need to develop processing units capable of performing high-speed, energy-efficient computing.
In response, both industry and academia are actively exploring alternative avenues from novel materials \cite{braga2009high,podzorov2004high}, architectures \cite{lecun2015deep,fan2004gpu}, to the investigation of new computational paradigms.

Among the emerging technologies, integrated photonics is a promising candidate for next-generation computation that can overcome the bottlenecks of their electrical counterparts. First, the speed of optical signals travel within optical waveguides surpasses that of electron-based transit through transistors with multiple fanouts by 1-2 orders of magnitude~\cite{gostimirovic2017ultracompact}. The delay and loss in waveguides are primarily determined by the optical path length. Additionally, a series of high-speed and energy-efficient (operating on the order of sub-picojoule per bit) devices for optical computing have been developed\cite{timurdogan2014ultralow,xu2021silicon,heni2019plasmonic}. Notably, the power consumption of transistor-based electrical circuits exhibits a cubic relationship with the clock frequency $f$\cite{mathew2005ghz}, whereas photonics-electronic platforms scale only linearly with $f$\cite{ying2020electronic}, effectively relaxing the frequency constraints associated with the power wall issue. Furthermore, as bosons, photons do not conform to the Pauli exclusion principle, allowing for the utilization of unique multiplexing techniques such as wavelength division multiplexing (WDM), which further increases the overall bandwidth. 
Moreover, compared to another photonic computing scheme in the form of free-space diffraction~\cite{zhou2021large, chen2023all}, integrated photonics offers superior compactness for higher-level integration. The advancement of silicon photonics has enabled the implementation of optical computing on low-cost PICs with high integration density, leveraging CMOS-compatible silicon manufacturing techniques.
As an increasing number of foundries develop their validated process design kits (PDKs), the integrated photonics industry is progressively moving towards standardization similar to that of the fabless semiconductor industry \cite{lim2013review}. This trend not only improves accessibility for designers and users but also offers more reliable performance.

Integrated photonics has emerged as a promising platform for AI accelerators, benefiting from its inherent attributes of high parallelism, low latency, and low power consumption.
In the last decade, a diverse range of PIC-based optical neural networks (ONNs) that implement multilayer perceptrons (MLPs)\cite{shen2017deep}, convolutional neural networks (CNNs)~\cite{chen2023all,meng2023compact,xu2021tops}, spiking neural networks (SNNs)~\cite{chakraborty2019photonic,feldmann2019all}, etc., have been reported, demonstrating remarkable performance on machine learning tasks. 
The fundamental operations of neural networks, involving data transfers and tensor operations, are achieved through the combination of passive optical devices and high-performance active photonic-electronic components\cite{shastri2021photonics,nozaki2019femtofarad}. Specifically, optical signals can be modulated by electrical signals and “multiplied” in accordance with the transmission function of the PIC. The hybrid photonic-electronic platform combines the adaptability of electronic control with the high-speed capabilities of optical computing. Recently, cutting-edge optical processing units have been reported with a matrix processing speed of 3.8 trillion operations per second (TOPS) via time-wavelength multiplexing\cite{xu2021tops}, while other works demonstrated ultra-low power consumption on the order of sub-femtojoules per bit\cite{heni2019plasmonic}.

While integrated photonics offers new opportunities, existing photonic-electronic computing systems still encounter several practical challenges, such as:
\begin{itemize}
    \item The typical micron-scale dimensions of optical elements in PICs are significantly larger than the transistors in cutting-edge VLSI technologies. Besides, a range of practical issues, such as footprint, control complexity, and accumulated loss, etc., limit the functionality and scalability of PICs for advanced computing applications.
    \item The widespread reliance on electrical components for electro-optical (E-O) modulation, parameter updates, data transfer, and analog-to-digital/digital-to-analog (A/D, D/A) conversion in photonic-electronic platforms leads to considerable energy consumption.
    \item The inherent challenges in PICs, such as training algorithms, on-chip implementation of nonlinearity for ONNs and system robustness against noise and crosstalk, require careful consideration in both hardware design and software coordination.
\end{itemize}

This review focuses on recent progress in photonic-electronic integrated circuits for computing.
Spanning from digital computing to analog AI accelerators, this paper is structured as follows. Section~\ref{section2} begins with an overview of the fundamental blocks in PIC-based digital computing, followed by a survey of recent highlights ranging from the implementation of logic gates to fully functional photonic processing units. 
Section~\ref{section3} focuses on the implementation of ONNs, covering aspects of photonic tensor cores, nonlinearity, and hardware-aware training strategies. Beyond a review from the device and circuit level, Sections~\ref{section4} and \ref{section5} provide a comprehensive analysis of recent photonic AI efforts from the perspectives of accelerator architectures and software-hardware co-design, respectively. 
The review culminates with Section~\ref{section6}, which offers an outlook on PIC-based optical computing and provides a summarizing conclusion.
\section{Survey of Optical Digital Computing with PICs}\label{section2}
PICs are comprised of a range of optical components, both passive and active, featuring various hardware implementations and circuit topologies to fulfill distinct functionalities. This section will focus on E-O digital logic, and provides a concise overview of recent progress in PIC-based digital computing, while highlighting these implementation techniques and associated challenges.
\subsection{Optical Logic Gates}
\begin{figure*}[htb]
  \centering
  \vspace{-5pt}
  \includegraphics[width=0.95\textwidth]{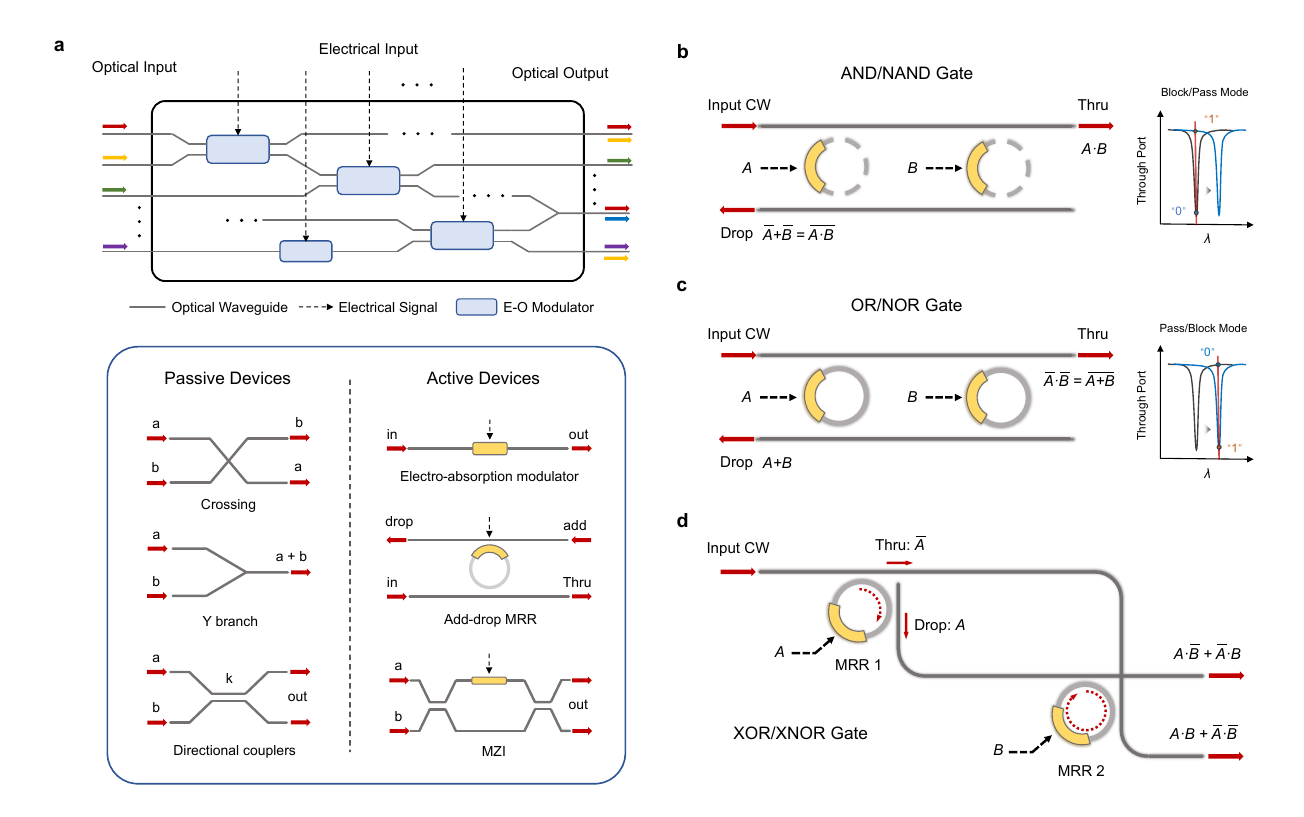}
  \vspace{-5pt}
  \caption{Implementations of electro-optic logic gates. (a) Schematic diagram of E-O logic composed of passive and active optical components. During each clock cycle, electrical inputs are used to configure the logic circuit, while light carries out logic operations based on the transmission characteristic of the logic block. (b)-(d) Schematic of MRR-based AND/NAND, OR/NOR, and XOR/XNOR gates, as proposed in Ref~\cite{tian2011proof,zhang2010demonstration}. In the diagram, the dotted line and the solid line represent MRRs functioning in the “block/pass” and “pass/block” modes, respectively. These configurations correspond to the outputs of `0' and `1' at the through port, given a logic `0' as the electrical input.}
  \vspace{-10pt}
  \label{LogicGate}
\end{figure*}

In the digital domain, both input and output are binary, and the resolution is defined by the number of bits and remains unaffected by the circuit size. A range of building blocks for optical digital computing on integrated photonic platforms, such as optical switches~\cite{jiang2019chip}, modulators~\cite{heni2019plasmonic,phare2015graphene}, interconnects and photodetectors~\cite{ying2015ultracompact,chaisakul2014integrated,ding2020ultra,michel2010high}, have been experimentally demonstrated. Basic logic operations (NOT, AND, OR, XOR, etc.), which are fundamental elements of digital systems, have been implemented by diverse PICs. Among these, electro-optic logic, also known as optical-directed logic, has been widely investigated by many research groups as well as foundries. As shown in Fig. \ref{LogicGate}.a, all E-O devices in the functional block, such as Mach-Zehnder Interferometers (MZIs), microring resonators (MRRs), microdisks, etc., are simultaneously configured by electrical signals. When light traverses the block, optical signals are modulated to execute logic operations in accordance with the PIC design and then propagated downstream or detected by monitors to read out the results. An important feature of the E-O digital logic is that each device is controlled by independent electrical input simultaneously and that signals are transmitted via light without the limitation of RC time constant and delay accumulations inherent in electrical systems. In other words, E-O logic merges the convenience and flexibility of electrical control with the high-speed capabilities of optical computing. Fig. \ref{LogicGate}.b-c show examples of E-O logic gates using two cascaded MRRs to perform 2-input AND/NAND and OR/NOR operations~\cite{tian2011proof}. The proposed optical logic gate leverages the transmission characteristic of add-drop MRRs, which work as optical switches to implement logic operations and generate complementary outputs at the through and drop ports. Additionally, by configuring the resonant states of MRRs with `0' operands via selecting continuous wave (CW) inputs at either on-resonance or off-resonance wavelengths, a single photonic circuit can execute a variety of logic functions. With consistent logic but varying PIC topologies, XOR and XNOR gates featuring a crossbar structure also have been demonstrated~\cite{zhang2010demonstration} (Fig. \ref{LogicGate}.d). Operating with similar mechanisms, MZI-based optical gates have also been extensively developed~\cite{kumar2014implementation,saharia2020comparative}.

In addition to E-O modulation, all-optical logic gate devices have also attracted attention. These devices have been experimentally demonstrated using various structures and phenomena, including photonic crystals~\cite{salmanpour2015photonic,hussein2018review}, surface plasmon polaritons (SPPs)~\cite{ota2016plasmonic}, nanowire networks~\cite{wei2011quantum}, and slot waveguides~\cite{pan2013optical,fu2012all}. A comprehensive classification, explanation of mechanisms, and comparative analysis of all-optical logic gates has been detailed in the prior review~\cite{jot2020all,fu2013silicon,anagha2022review}. Compared to E-O logic, all-optical logic offers the potential for higher operation speed and bandwidth without extra energy consumption associated with O-E-O conversion. However, the implementation of all-optical logic confronts several practical challenges. Firstly, the complexity and requirements of design and fabrication (e.g., the transmission characteristic of photonic crystal is highly sensitive to its lattice constant) lead to inherent instability and a low contrast ratio between logical states~\cite{fu2012all}. Secondly, while all-optical gates can be more energy-efficient in signal processing, they often require higher optical signal power (mW-level) to compensate for higher losses or to induce the necessary nonlinear effects for switching. Furthermore, the limited functionality and scalability of all-optical logic, coupled with its higher cost, restrict its widespread application compared to E-O logic.

In recent years, integrated photonics have expanded beyond traditional classical optics, emerging as a compelling platform for quantum information science. Quantum logic gates based on the aforementioned active/passive devices have been widely reported \cite{politi2008silica,fedorov2012implementation,crespi2011integrated}. Quantum PICs offer several significant advantages over bulk optics in the realm of quantum computing. First, PICs enable precise control of phase, polarization, and spatial mode with higher stability, which are essential for manipulating quantum states. Second, silicon exhibits a high third-order nonlinear coefficient $\chi^{(3)}$, facilitates the effective implementation of on-chip single/entangled photon sources through optical processes such as four-wave mixing (FWM)~\cite{harris2016large}. Third, PICs can integrate the fundamental building blocks of quantum computing—such as photon source, modulators, and single-photon detectors, etc., —in monolithic, hybrid, or heterogeneous configurations \cite{elshaari2020hybrid}. This integration yields scalable, robust, and reconfigurable circuits capable of handling complex quantum computing tasks~\cite{wang2020integrated,harris2016large,pelucchi2022potential}.

\subsection{Combinational Logic and Reconfigurable PICs} \label{subsec:ReconfigurablePIC}
 In digital circuits, the output of a combinational logic unit is determined solely by the current input combination, without dependence on previous states. Similarly, the implementation of optical combinational logic could begin with extracting the logical expression from its truth table, followed by designing the corresponding PIC based on the simplified expression. The implementation can rely on assembling fundamental logic gates or leveraging the unique characteristics of optical components or multiplexing techniques for fewer devices and compact layouts. Employing these strategies, diverse optical combinational logic units, including but not limited to adders~\cite{ying2018silicon}, comparators~\cite{yang2015demonstration}, encoders~\cite{tian2011demonstration, xiao2018experimental} and decoders~\cite{mehdizadeh2016novel}, have been reported.

\begin{figure}[htb]
  \centering
  \vspace{-5pt}
  \includegraphics[width=0.46\textwidth]{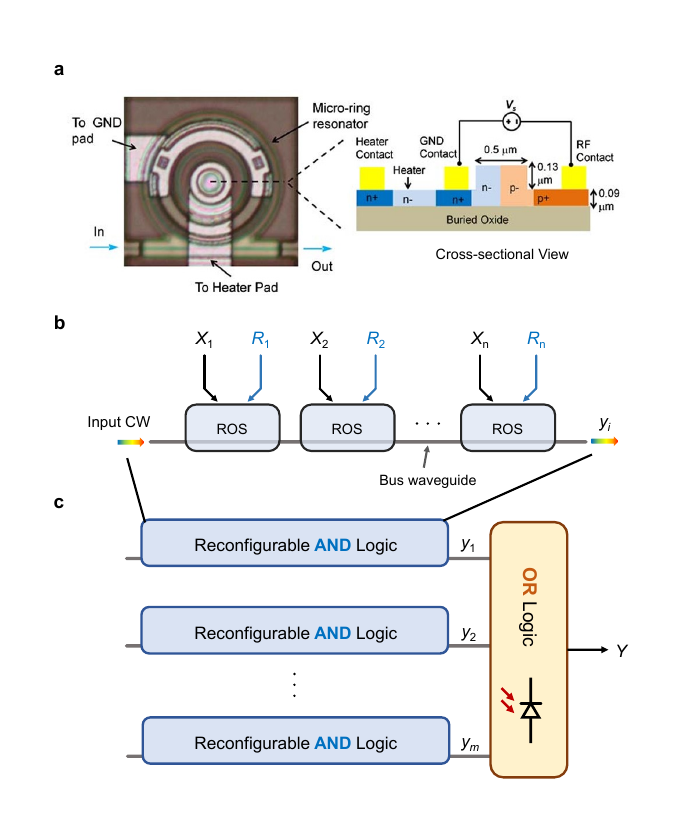}
  \vspace{-5pt}
  \caption{Reconfigurable PICs for arbitrary combinational logic. (a). Optical micrograph and cross-sectional diagram of a reconfigurable MRR featuring two modulation mechanisms. The \textit{p-i-n} junction is applied by RF signal for input encoding, while the microheater is connected to the low-speed DC signal for resonance mode reconfiguration~\cite{qiu2014reconfigurable}. (b)-(c) Schematic of a PIC architecture that  enables the implementation of arbitrary combinational logic expressions based on reconfigurable optical switchs (ROSs).}
  \vspace{-5pt}
  \label{ReconfigurablePIC}
\end{figure}

 The aforementioned optical logic gates and units, tailored for specific tasks, are constrained by a fixed or limited logic representation. The inherent limitation not only complicates the development process but also increases the cost. To address this challenge, reconfigurable PICs offer a promising solution by programming the operational states of optical switches within a pre-designed framework using additional signals. The input operands and the reconfiguration signal can be independently managed by two separate control units within the modulator, such as the dual arms of MZIs. Generally, reconfiguration signals do not require high-speed modulation to the same extent as the input signals used as logic operands.  Qiu et al. proposed a reconfigurable logic unit based on MRRs embedded with two modulation mechanisms~\cite{qiu2014reconfigurable}. As is shown in Fig. \ref{ReconfigurablePIC}.a, the logic operand is modulated at high speeds via the \textit{p-i-n} junction operating in the carrier injection mode, while the resonant state of MRR can be reconfigured using the microheater. Additionally, multi-operand modulators and non-volatile devices are also promising candidates for reconfigurable PICs~\cite{ruhul2024reconfigurable,ying2019integrated,fang2021non}.

An arbitrary combinational logic expression $Y$ with $n$ inputs $X_1, X_2, \ldots, X_n$ can be represented as a sum of products derived from these inputs, which can be expressed as:
\begin{equation}\label{CombinationalLogic}
\vspace{-5pt}
Y = y_1 + y_2 + \ldots + y_m, \text{ where } y_i = \prod_{k=1}^{n} X_k \text{ or } \overline{X_k}
\vspace{-3pt}
\end{equation}
In this expression, $\overline{X_k}$ denotes the complement logic of input $X_k$. Using reconfigurable optical switches, the architecture illustrated in Fig. \ref{ReconfigurablePIC}.b-c theoretically can implement arbitrary logic functions conforming to the expression format presented in Eq. (\ref{CombinationalLogic}). The product term $y_i$, i.e., the logic AND operation, can be implemented by $n$ serially connected reconfigurable optical switches along single bus waveguides. This block yields logic `1' output only when all switches are in the ``pass" state. Reconfiguration signals $R_i$ are used to determine whether the corresponding operand contributes complement logic to $y_i$. In contrast to electrical digital computing, the OR operation for optical signals can be directly implemented using a combiner and detected by a photodetector. It is important to recognize that when product terms are represented by the same wavelength, the amalgamation may result in logical errors due to coherent interference. This issue can be avoided either by using distinct wavelengths for each branch~\cite{qiu2014reconfigurable}, or equipping each branch with photodetectors individually~\cite{qi2019silicon}. Besides, given that photodetectors operate as current sources, a straightforward parallel configuration could achieve electrical OR logic without introducing additional delays. 
While this strategy offers a general solution for optical combinational logic, the limited scalability restricts its practical applicability in scenarios involving numerous operands. For an $n$-operands system, the complexity of this processing unit escalates as $O(n \cdot 2^n)$, indicating an exponential increase in the requisite number of switches with $n$. Furthermore, issues related to power consumption and accumulated losses also need to be considered. The dynamic power of the system $P_{\text{dynamic}}$ can be expressed as:
\begin{equation}\label{P_dynamic}
P_{\text{dynamic}} = \frac{1}{4} \alpha CV^2 f
\end{equation}
where, $\alpha$ is the activity factor and $C$ is the total capacitance of the E-O modulators,  which increases proportionally with the number of modulators. It is important to note that while Eq. (\ref{P_dynamic}) has a similar expression to that used for CMOS transistors, the supply voltage $V$ does not necessarily scale with $f$. Therefore, to reduce $P_{\text{dynamic}}$, two straightforward strategies can be considered: (1) Decrease the number of modulators, for example, by employing multi-operand logic gates to squeeze logic functions into fewer devices \cite{ying2019integrated}; (2) Utilize devices with low capacitance, such as microdisks with capacitance in 10's fF \cite{ying2020electronic}. Additionally, when present, thermal tuning typically dominates the power consumption for modulation. This portion of power consumption can be reduced or even eliminated through device-level optimization, such as post-fabrication trimming \cite{atabaki2013accurate}, or using energy-efficient tuning mechanisms (discussed in Section \ref{ProgrammableModulation}). The total power consumption also includes the laser and detection parts, as well as the potential static power consumption. The laser power is primarily determined by the losses, typically at the mW level with dB/cm-level loss, which also correlates with the PIC scale. 

In the realm of electrical digital design, electronic design automation (EDA) tools assist designers by automatically generating and optimizing circuits from logic expressions, as well as auto-placing and routing in physical design. With the advancement of PIC-based computing, logic synthesis methods specified for PIC design have been proposed, enabling large-scale design automation and optimization of compact PICs for digital computing~\cite{testa2018logic,ying2018automated,zhao2019exploiting}.
\subsection{Toward a Fully-functional EPALU}
\begin{figure*}[htbp]
  \centering
  \vspace{-10pt}
  \includegraphics[width=1\textwidth]{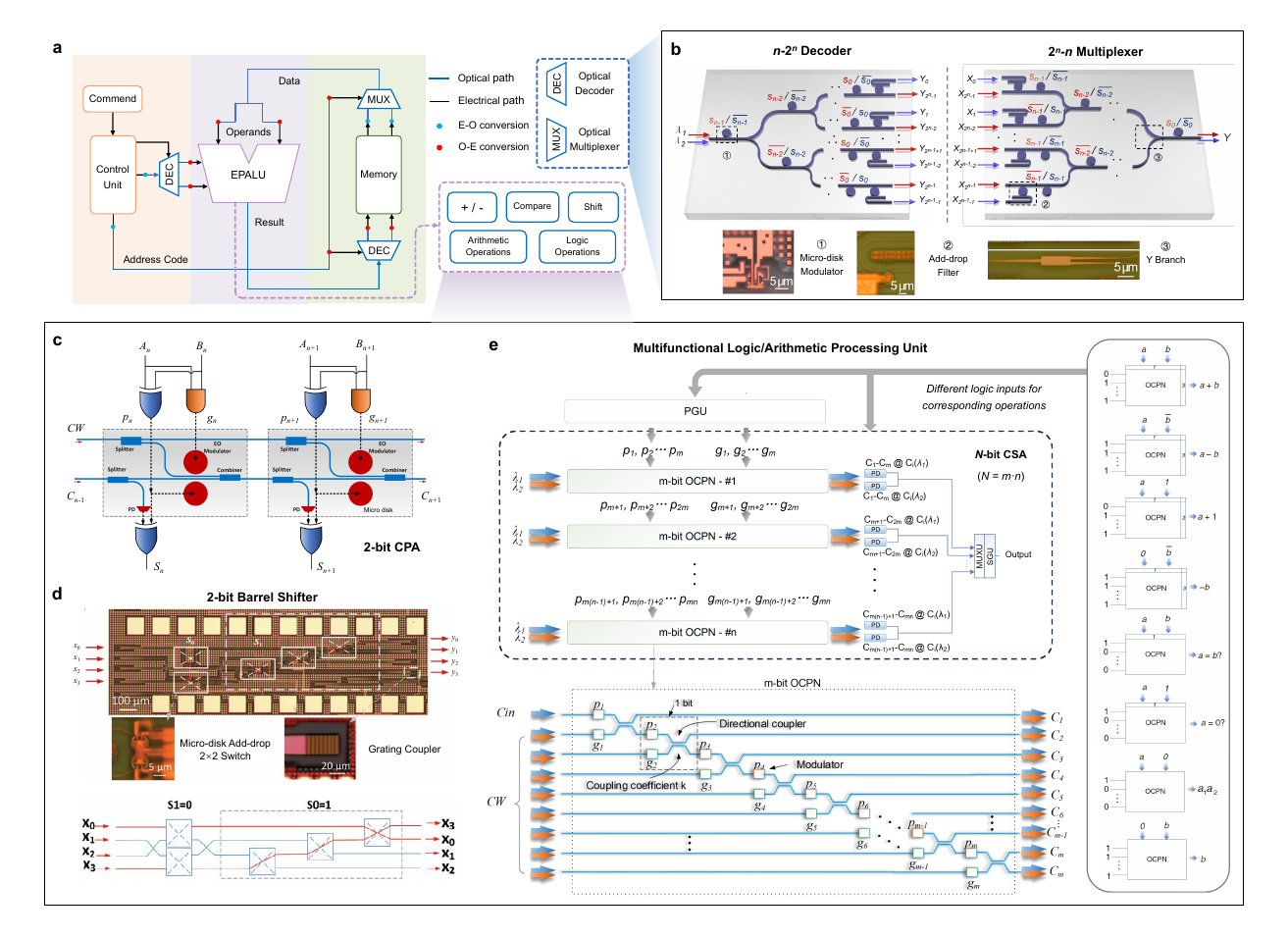}
  \vspace{-10pt}
  \caption{EPALU architecture for high-performance digital computing. (a) Schematic of electronic–photonic microprocessor with its building blocks and data path. (b) Schematic of an $n$-$2^n$ E-O decoder and $2^n$-$n$ O-E multiplexer~\cite{feng2020wavelength}, where $s_i$ refers to the electrical input signal. (c) Schematic of a 2-bit CPA using E-O logic~\cite{ying2018silicon}. (d) Layout of a 2-bit barrel shifter using microdisk add-drop switch array, and the optical datapath with $S = \text{`01'}$~\cite{feng2022integrated}. (e) The architecture of the WDM-based $N$-bit multifunctional processing unit consists of a $(p, g)$ generation unit (PGU), $n$ sets of $m$-bit optical carry propagation networks (OCPNs), and an array of photodetectors (PDs) along with a network of electronic multiplexer units (MUXU) and an electronic sum generation unit (SGU)~\cite{ying2020electronic}. With different input combinations, EPALU can perform various logic/arithmetic functions (right). }
  \vspace{-10pt}
  \label{EPALU}
\end{figure*}

Besides the aforementioned advantages of optical logic units, unique multiplexing techniques play a pivotal role in further improving computing capacity and performance. An example is the WDM-based electronics-photonic arithmetic logic unit (EPALU) (Fig. \ref{EPALU}.a)~\cite{ying2020electronic}. The arithmetic logic unit (ALU), which performs arithmetic and bitwise operations, is an essential component of modern computing systems. Within ALUs, the full adder plays a critical role, with its logic being expressed as follows:
\begin{equation}\label{adder}
\begin{split}
C_n = p_n \cdot C_{n-1} + &g_n \; , \quad S_n = C_{n-1} \oplus p_n \\[5pt]
g_n = A_n \cdot B_n \; &, \quad p_n = A_n \oplus B_n
\end{split}
\end{equation}
where $A$, $B$, and $C$ represent the two operands and carry, respectively; $p$ and $g$ denote propagation and generation, and the subscripts indicate the specific bits. Based on these expressions, the scalable electro-optic carry propagation adder (CPA) has been developed~\cite{ying2018silicon}, with the schematic shown in Fig. \ref{EPALU}.c. For a $N$-bit CPA, the carry output from one stage is the carry input to the next stage, thus the final result cannot be calculated until the carry has rippled through all stages. As an optimized architecture, the carry select adder (CSA) splits $N$-bit operands into $n$ $m$-bit CPA ($N = m \times n$). It speeds up addition by computing two possible outcomes for each $m$-bit CPA simultaneously—assuming carry-in values of `0' and `1'—and then selects the correct result based on the actual carry-in using multiplexers (MUXs)~\cite{sklansky1960conditional}. As a trade-off, CSAs require two sets of circuits with different carry inputs. However, in the optical domain, this can be efficiently implemented with a single optical path in the PIC, where different carry signals are encoded into two distinct wavelengths using WDM (Fig. \ref{EPALU}.e). Beyond arithmetic addition, Ying et al. developed that the multifunctional architecture could perform addition, subtraction, comparison, and bitwise operations operating at 20 GB/s with various input combinations \cite{ying2020electronic}. Based on time-space multiplexing, Zhang et al. demonstrated a photonic-electronic digital multiplier capable of processing up to 32 $\times$ 4-bit binary inputs at 25 Mbit/s~\cite{zhang2024time}. Additionally, the E-O shifter within the EPALU architecture has also been experimentally demonstrated (Fig. \ref{EPALU}.d)~\cite{feng2022integrated}.

From the perspective of computer architecture, the ALU operates under the directives issued by electronic control units, with its inputs being fetched from memory. Upon completing a designated operation, the ALU's output is then stored back in memory to be accessed for subsequent computations (Fig. \ref{EPALU}.a). Although optical computing has significantly reduced the latency of arithmetic operations, data access and E-O/O-E conversions can create major bandwidth bottlenecks and serve as a significant source of energy consumption in the computing system. Therefore, the exploration of high-speed interconnects between different modules within an electronic–photonic microprocessor is important as well~\cite{sun2015single,feng2020wavelength}. In Ref~\cite{feng2020wavelength}, a decoder and multiplexer designed for the EPALU architecture have been demonstrated, achieving data transportation and processing at a speed of 20 Gb/s (Fig. \ref{EPALU}.b). Another potential approach to reduce the time and energy costs of E-O/O-E conversion is through the implementation of various forms of optical memory, as discussed in Ref~\cite{alexoudi2020optical,mcmahon2023physics}.

\section{PIC-Based Analog computing for AI: Fundamentals and Implementation}\label{section3}
Undoubtedly, modern AI, functioning on digital computing systems, has achieved significant progress in diverse fields and has even exceeded human performance in specific tasks. With the advancements in integrated photonics, the PIC platform emerges as a compelling candidate for AI accelerators. Leveraging the optical logic gates and computing units detailed in Section~\ref{section2}, some studies have illustrated the capability of PICs to perform tensor operations, including accumulation, dot products, and matrix-vector multiplications (MVMs), for diverse machine learning applications within the digital domain~\cite{liu2019holylight,zheng2023priml,zokaee2020lightbulb}. These studies have highlighted the inherent advantages of optical computing in terms of latency, speed, and power efficiency. However, the digital representation can encounter challenges stemming from hardware complexity overhead and speed reduction caused by the sampling and digitization into binary streams processed by logic units. These challenges are especially significant in the context of high-throughput or high-precision tensor operations for various machine-learning tasks.

\begin{figure}[htb]
  \centering
  \vspace{-5pt}
  \includegraphics[width=0.45\textwidth]{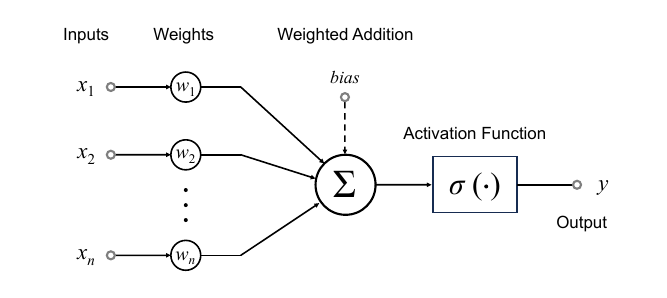}
  \vspace{-5pt}
  \caption{Schematic of an artificial neuron with simple synaptic model.}
  \vspace{-10pt}
  \label{Neuron}
\end{figure}

On the other hand, the human brain, operating as an analog signal ``processor", is estimated to perform at a rate of $10^{18}$ multiply-accumulate (MAC)/s with a power consumption estimated at just $\sim$10-20 W~\cite{chen2021energy,hsu2014ibm}, demonstrating remarkable efficiency compared to the substantial energy requirements of cutting-edge AI. This efficiency can potentially be attributed to the parallel processing capability and reduced precision requirements inherent in analog computing. While the mechanisms of brain function remain incompletely understood, there is growing interest in incorporating analog computing into machine learning. Before discussing the details of implementing optical analog computing for AI, it is helpful to provide a concise overview of artificial neural networks (ANNs) and the neuron model. The schematic of an artificial neuron with a basic synaptic model is illustrated in Fig. \ref{Neuron}, where \textit{\textbf{x}}, \textit{\textbf{y}}, and \textit{\textbf{w}} represent the inputs from the pre-synaptic neuron, post-synaptic output, and weights of the connection, respectively. The activation functions $\sigma(\cdot)$, such as sigmoid, ReLU, and the leaky integrate-and-fire (LIF) function for SNNs, introduce nonlinearity into the model along with various engineering considerations~\cite{dubey2022activation,teka2014neuronal}. 

\subsection{Programmable modulation for optical analog computing} \label{ProgrammableModulation}
In analog AI accelerators, both inputs \textit{\textbf{x}} and weights \textit{\textbf{w}} could correspond to a higher resolution, in contrast to the binary values in digital computing circuits. The hardware implementation of the above process using PICs requires the reconfigurable programming of network parameters, which relies on the modulation of optical components. While the weights in a trained model may remain fixed during the inference process, it is still necessary to calibrate the network through modulation due to the fabrication variations of PICs. A number of modulation mechanisms have been developed, among which tuning the effective refractive index $n_{\text{eff}}$ of waveguides is a widely adopted approach in ONNs.

\subsubsection{Thermal tuning mechanism}
Based on thermo-optic effects, the transmission characteristics of devices can be modulated by changing the $n_{\text{eff}}$ of waveguide through integrated filament microheaters. The heat generated by the ohmic microheater is proportional to the square of the bias voltage. Fig. \ref{ModulateTech}.a shows the schematic and transmission curve of a thermo-optic MZI fabricated by \textit{Advanced Micro Foundry}~\cite{feng2022compact}. Thermal tuning demonstrates the adaptability to various devices and substrate materials, with minimal constraints imposed by the fabrication process. Furthermore, compared with other mechanisms (especially for silicon), it can induce large changes in $n_{\text{eff}}$ within a small footprint. However, thermo-optic devices encounter difficulties in achieving high-speed modulation, currently limited to a few hundred KHz~\cite{harris2014efficient}, although sufficient for inference tasks with static weights. Additionally, thermal tuning is a volatile configuration process, requiring continuous external biasing and power supply (typically $\sim$mW level) to hold its functionality. Another issue arises from thermal crosstalk when heat dispersion cannot be adequately contained without physical constraints, such as trenches, which need appropriate consideration in both schematic and layout design.

\subsubsection{Field-effect tuning mechanism}
Except for the thermo-optic effect, the $n_{\text{eff}}$ can also be tuned by electric fields. In silicon photonics, a straightforward approach involves doping the silicon waveguide and applying an external electric field to manipulate the carrier concentration and tune $n_{\text{eff}}$. A variety of modulators operating in carrier-injection (forward-bias \textit{p-i-n}), carrier-depletion (reverse-bias \textit{p-n} junction), and carrier-accumulation (metal-oxide-semiconductor capacitor, MOSCAP) mode have been widely demonstrated (Fig. \ref{ModulateTech}.b)~\cite{manipatruni2010ultra,green2007ultra,moss20131,xu2013high,wang2013optimization,zhang2023harnessing}. These CMOS-compatible mechanisms allow gigahertz-level tuning speeds, making them suitable for high-speed encoding in ONNs. Particularly, in depletion mode, the bandwidth is determined by the majority carriers' dynamics, which are not limited by the slower processes of carrier generation and recombination~\cite{xu2013high}. Moreover, while the depletion-mode modulator remains a volatile device, it exhibits low static power consumption attributed to the reverse-biased junction. 
Compared to thermo-optic devices, the free-carrier effect-based modulators typically require longer modulation lengths and larger footprints, primarily due to their lower tunning efficiencies (as evaluated by voltage-length product $V_{\pi}L$) or the necessity for traveling-wave electrodes for high-speed modulation. 
Expanding beyond the conventional silicon platform, an alternative approach is the utilization of materials with significant electro-optic effects—such as the free-carrier plasma dispersions, 
the Pockels effect, the Kerr effect, and the Quantum Confined Stark Effect (QCSE)—for the core or cladding of waveguides. The modulation efficiency of these devices is intrinsically linked to the material properties. Prominent examples include III–V semiconductors~\cite{park2018ingaasp,zhang2023harnessing,hiraki2017heterogeneously}, lithium niobate~\cite{wang2018integrated}, and some polymers~\cite{zhang2012highly,zhang2016high}. A comprehensive discussion of all these mechanisms exceeds the scope of this review. Interested readers are referred to relevant reviews and reports for further information \cite{liu2015review,sinatkas2021electro,hsu2021mos}.

Most modulators discussed above have nonlinear transmission curves, primarily due to the E-O modulation mechanisms involved. Although the Pockels effect provides a linear change in $\Delta n$ in response to electric field intensity, the inherent transfer function of the modulator itself can still exhibit a nonlinearity (such as the sinusoidal response of MZI). To reduce undesired nonlinearity in computing architectures, several solutions have been proposed: (1) Introduce a nonlinear mapping during D/A conversion, although this may necessitate additional data processing; (2) Utilize the nearly linear segment of the transmission curve, at the tradeoff of a reduced modulation dynamic range; (3) Design PICs that compensate for nonlinearity by novel devices or modulation mechanisms~\cite{xiong2014linear,yamazaki2016optical,zhang2012highly,lee2009bias,feng2022ultra}. Additionally, the impact of nonlinearity on system noise also needs consideration. Within modulators exhibiting strong nonlinearity and steep slopes, such as high-Q MRRs, slight noise or offsets in the drive signal can potentially result in significant deviations.

\subsubsection{Non-volatile modulation}
\begin{figure*}[htbp]
  \centering
  \vspace{-5pt}
  \includegraphics[width=0.96\textwidth]{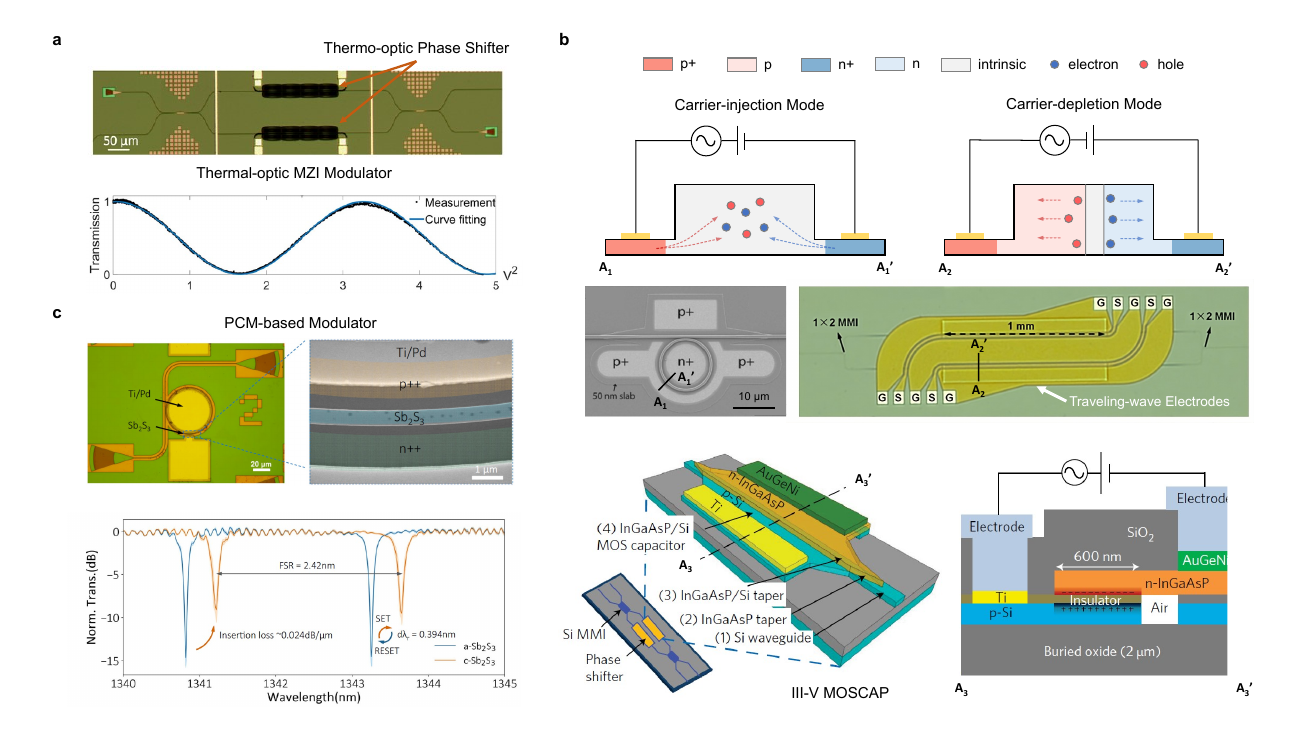}
  \vspace{-5pt}
  \caption{Modulation techniques of photonic-electronic devices on PICs. (a) Optical micrograph and the normalized transmission curve of a thermo-optic MZI. Each arm of the MZI incorporates a microheater as a phase shifter~\cite{feng2022compact}. While both arms can be used to modulate the output signal, typically, one is used to set the modulator output at the quadrature point for a high extinction ratio in monotonic modulation. (b) Schematics and micrographs of free-carrier effect-based modulators working in carrier-injection mode~\cite{chen2009integrated}, carrier-depletion mode~\cite{wang2013optimization}, and by the MOSCAP-driven tuning mechanism~\cite{hiraki2017heterogeneously}. (c) Top, optical micrograph of an MRR modulator loaded with 10 \(\mu\)m long 20-nm-thick Sb\textsubscript{2}S\textsubscript{3} and doped silicon PIN heater. Bottom, normalized transmission spectra of MRR when switching between two phases of Sb\textsubscript{2}S\textsubscript{3}~\cite{chen2023non}.}
  \vspace{-10pt}
  \label{ModulateTech}
\end{figure*}

Each aforementioned mechanism is a volatile process and requires continuous power supply even for infrequent or static programming tasks. Non-volatile phase change materials (PCMs) offer a unique opportunity to avoid these scenarios. PCMs have two switchable phases, i.e. amorphous and crystalline states, with drastically different $n_{\text{eff}}$, and can achieve reversible phase transitions within various temperature ranges. For the PIC platform, heating can be achieved by an integrated pulse microheater or alternatively, through light signals themselves, enabling 'all-optical' modulation. Chen et al. developed several non-volatile silicon photonics modulators with Sb\textsubscript{2}S\textsubscript{3} PCM cladding and experimentally demonstrated 5-bit multilevel programming with a high cyclability $>$1600 switching events (Fig. \ref{ModulateTech}.c)~\cite{chen2023non}. Besides, all-optical SNNs using chalcogenide PCM GeSbTe (GST) also have been reported~\cite{feldmann2019all}, and more details will be discussed in Section.~\ref{AllOptical}. A brief summary of modulation techniques with Si waveguide is shown in Table~\ref{tab:modulation_techniques}.
\renewcommand{\arraystretch}{1.25}
\begin{table*}[htb]
    \centering
        \begin{threeparttable}
        \caption{Summary of Modulation Techniques with Silicon Waveguide} 
        \label{tab:modulation_techniques} 
        \begin{tabular}{@{}cccC{1.5cm}C{1.5cm}C{2.8cm}C{1.8cm}cc@{}} 
        \toprule
        Mechanism         & $V_\pi L$ (V$\cdot$cm) & IL (dB) & $f_{3\text{dB}}$ & Data Rate& Energy Efficiency       & Active Length     & Modulator  & Ref  \\ 
        \midrule
        Thermo-optics     & $\sim$0.023    & $<$1   & $\sim$4 KHz & —        & $\sim$1.36 mW*              & $\sim$170 $\mu$m & MZM        & \cite{feng2022compact} \\
        Carrier Injection & 0.036           & 12      & 16 GHz  & 10 Gb/s      & 51 mW* / 5 pJ/bit            & 200  $\mu$m       & MZM        & \cite{green2007ultra} \\
        Carrier Depletion & 1.62–2.05       & 3.9     & 30 GHz  & 44  Gb/s     & 2.84 pJ/bit                  & 1000 $\mu$m       & MZM        & \cite{wang2013optimization} \\
        Si-MOSCAP         & $\sim$0.8      & 9       & 50 GHz  & $>$100 Gb/s  & —                            & R = 15 $\mu$m     & MRR        & \cite{zhang2023harnessing} \\
        III-V-MOSCAP      & 0.09            & 1       & 2.2 GHz & 32 Gb/s      & —                            & 250 $\mu$m        & MZM        & \cite{hiraki2017heterogeneously} \\
        Graphene-MOSCAP   & 0.28            & $\sim$7 & 5 GHz   & 10 Gb/s      & 1 pJ/bit                     & 300 $\mu$m        & MZM        & \cite{sorianello2018graphene} \\
        PCM               & $L_\pi$=30.7$\mu$m    & $<$1    & $\sim$10 KHz  & —      & \makecell[c]{11.6 mJ for SET 
                                                                                 \\ 197 nJ for RESET}         & R = 30 $\mu$m     & MZM/MRR    & \cite{chen2023non} \\ 
        \bottomrule
        \end{tabular}
        \begin{tablenotes}
            \item[] * Static power consumption
            \item[] $\sim$ Measured from figures or calculated based on other parameters in the articles
            \item[] - Not mentioned or unavailable in the article
        \end{tablenotes}
    \end{threeparttable}
    \vspace{-10pt}
    \label{Table1}
\end{table*}
\subsection{Implementations of Photonic Tensor Cores}
As illustrated in Fig.~\ref{Neuron}, the interconnections in ANNs can be conceptualized as weights akin to synaptic coupling coefficients in biological systems. This analogy extends to representing these connections through tensor operations, thereby abstracting the complex interactions in a computationally manageable form. Building upon the aforementioned modulation mechanisms, various active devices and encoding mechanisms have been utilized effectively in photonic neurons. This section categorizes ONNs from multiple perspectives, which aims to provide a comprehensive analysis and comparison of the implementations of photonic tensor cores (PTCs) from diverse optical components to the architectural design.

\subsubsection{Coherent vs incoherent ONN}
\begin{figure*}[htb]
  \centering
  \vspace{-5pt}
  \includegraphics[width=1\textwidth]{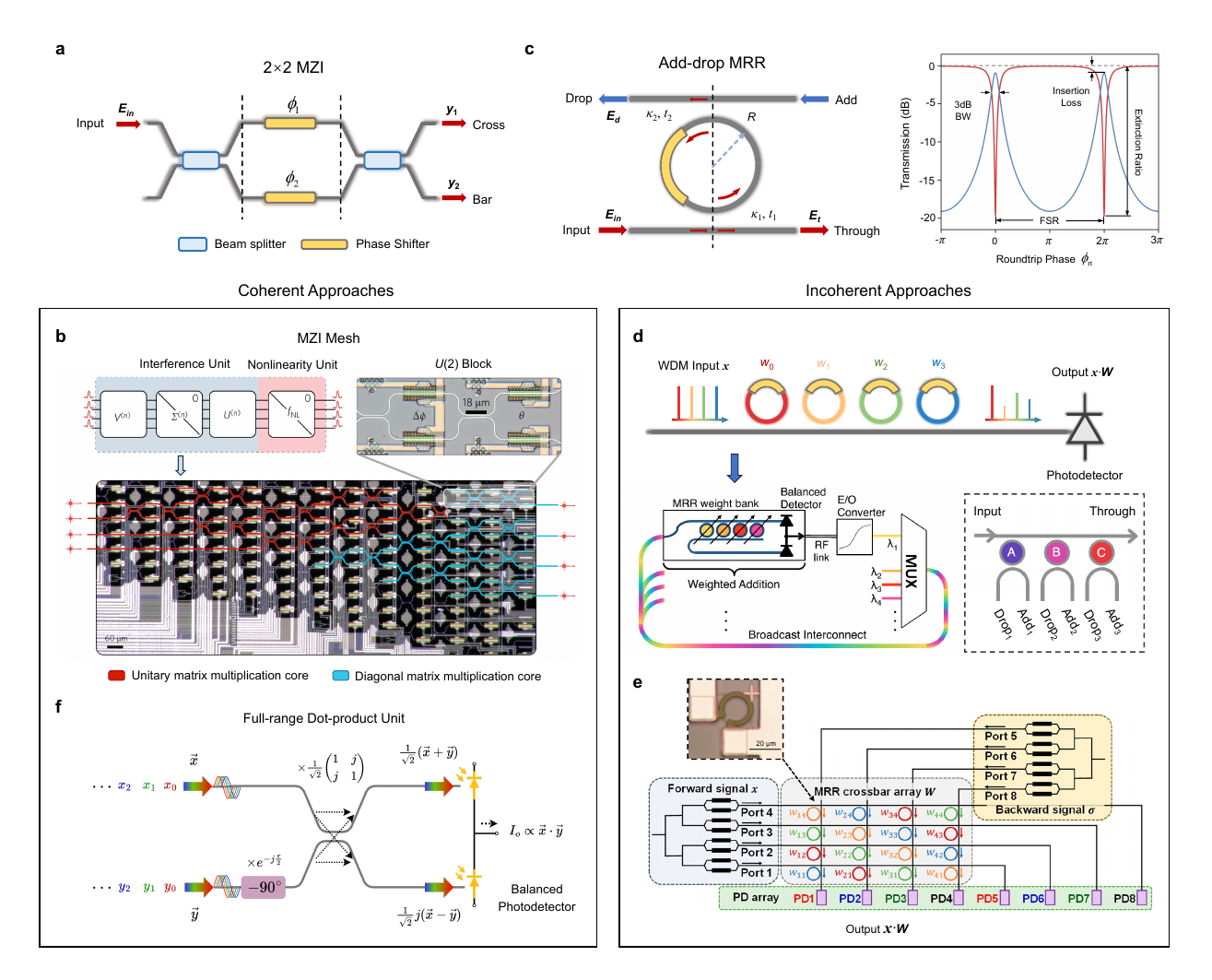}
  \vspace{-5pt}
  \caption{Various implementations of photonic-electronic tensor cores. (a) Schematic of a 2$\times$2 MZI where the beam splitters can be implemented by multi-mode interferometer (MMIs) or directional couplers. An alternative configuration of MZI features a single input and output, with the splitter designed as a Y branch. (b) An MZI-based coherent photonics tensor core for ONNs~\cite{shen2017deep}. The weight matrix can be represented by SVD in the form $\bm{U\Sigma V^{\dagger}}$. Due to chip size and complexity considerations, this work only implements $\bm{U}$ and $\bm{\Sigma}$ on a single pass through the circuit. (c) Schematic and power spectral responses of an add-drop MRR with power coupling coefficients ${\kappa_1}^2 = {\kappa_2}^2 = 0.2$ (assuming the coupling junctions are lossless, i.e., $\kappa^2 + t^2 = 1$) and 5\% round-trip power loss. (d) The weight bank architecture utilizes MRRs as tunable filters for parallel modulation of WDM signals~\cite{tait2016microring}. The series-connected MRRs share a single bus waveguide as a through port, while the other waveguide can be shared or independent, enabling the selective addition or removal of wavelengths. (e) Schematic of 4$\times$4 MRR crossbar array demonstrated in Ref~\cite{ohno2022si}. Input signals are first modulated via MZMs, and then distributed and weighted through the crossbar array, followed by a photodetector array that sums the weighted WDM elements. (f) The Schematic of an optical tensor core supports dynamic full-range general matrix multiplication based on WDM and coherent interference~\cite{zhu2023dota}.}
  \vspace{-10pt}
  \label{SynapseImplement}
\end{figure*}

From the viewpoint of signal properties, ONNs can be classified into coherent and incoherent systems. Within coherent ONNs, both weights and inputs can be encoded in the complex plane, allowing multiplication through lossless interference. For instance, a pair of beam splitters and phase shifters in the form of MZIs are widely adopted for conducting linear operations in coherent ONNs due to the programmable amplitude and phase response~\cite{reck1994experimental}. The schematic of a 2$\times$2 MZI is shown in Fig. \ref{SynapseImplement}.a. Assuming that each beam splitter is an ideal 50:50 directional coupler (i.e., with a coupling coefficient $\kappa = 1/\sqrt{2}$), the transition matrix of the MZI can be represented as:
\begin{equation}\label{MZI_TransMatrix}
T_{\text{MZI}} = \frac{1}{\sqrt{2}}
\begin{bmatrix}
1 & j \\[4pt]
j & 1
\end{bmatrix} \cdot
\begin{bmatrix}
e^{-j\phi_1} & 0 \\[4pt]
0 & e^{-j\phi_2}
\end{bmatrix} \cdot \frac{1}{\sqrt{2}}
\begin{bmatrix}
1 & j \\[4pt]
j & 1
\end{bmatrix}
\end{equation}
Here, $\phi_1$ and $\phi_2$ represent the phase changes as signals traverse two arms. In some cases, $\phi_1$ and $\phi_2$ are denoted as $\Delta\phi = \phi_1 - \phi_2$ and 0, respectively, given that the response is predominantly governed by the phase difference between its two arms. If only one input port receives a signal $\bm{E}_\textbf{in}$, the field and power intensities at the output port can be expressed as follows:
\begin{equation}\label{MZI_TransFuc1}
\begin{split}
y_1 &= -j\cdot \exp\left[-j\left(\phi_2 + \frac{\Delta\phi}{2}\right)\right] \cdot \sin\left(\frac{\Delta\phi}{2}\right) \cdot E_{\text{in}} \;; \\[8pt]
y_2 &= \phantom{-}j\cdot \exp\left[-j\left(\phi_2 + \frac{\Delta\phi}{2}\right)\right] \cdot \cos\left(\frac{\Delta\phi}{2}\right) \cdot E_{\text{in}}
\end{split}
\end{equation}
\begin{equation}\label{MZI_TransFuc2}
|y_{1,2}|^2 = |E_{\text{in}}|^2 \cdot \frac{1\mp\cos\Delta\phi}{2}
\end{equation}
This implies that the phase and amplitude of output can be modulated by varying $\Delta\phi$ from 0 to $\pi$. The smooth transmission curve, stemming from the cosine term, enables high-resolution analog computing and enhances noise resistance. In a notable case, the two arms are encoded by opposite phase shifts, represented as $\phi \pm \Delta\phi/2$ for $\phi_1$ and $\phi_2$, respectively. When incorporated into Eq. (\ref{MZI_TransMatrix}), the Mach-Zehnder modulator (MZM) working in “push-pull” mode allows for the independent modulation of field intensity while maintaining a constant phase. 

An MZI cascaded with a phase shifter can implement a 2$\times$2 unitary transformation:
\begin{align}\label{UnitaryMZI}
U(2) &= \overbrace{\left[ \begin{array}{cc}
e^{-j\theta} & 0 \\[4pt]
0 & 1 \\
\end{array} \right]}^{\text{Phase Shifter}}
\cdot
\overbrace{\frac{1}{2}\left[ \begin{array}{cc}
e^{-j\Delta\phi}-1 & j(e^{-j\Delta\phi}+1) \\[6pt]
j(e^{-j\Delta\phi}+1) & -(e^{-j\Delta\phi}-1) \\
\end{array} \right]}^{\text{MZI}} \nonumber \\[8pt]
&= \frac{1}{2}\left[ \begin{array}{cc}
e^{-j\theta}(e^{-j\Delta\phi}-1) & je^{-j\theta}(e^{-j\Delta\phi}+1) \\[6pt]
j(e^{-j\Delta\phi}+1) & -(e^{-j\Delta\phi}-1) \\
\end{array} \right]
\end{align} 
Unitary matrices of rank $N$ can be decomposed into sets of $U(2)$ blocks, which can be implemented using cascaded 2$\times$2 MZIs to form a mesh structure. Despite the unitary nature of the transition matrix in an MZI-mesh, an arbitrary weight matrix for an ONN can be implemented by singular vector decomposition (SVD). The SVD approach was thoroughly discussed by Miller et al. in 2003 and first experimentally implemented in ONNs by Shen et al. as shown in Fig. \ref{SynapseImplement}.b~\cite{miller2013self,shen2017deep}. Specifically, an arbitrary real-valued matrix $\bm{M}$ can be decomposed into the form $\bm{U\Sigma V^{\dagger}}$, where $\bm{\Sigma}$ is a diagonal matrix and the remaining two are unitary matrices, which can be implemented with a set of tunable attenuators and an MZI mesh, respectively. 

Another characteristic of the devices based on phase modulation is their broad spectral bandwidth. During the modulation process, the phase change $\psi$ can be expressed as follows:
\begin{equation}\label{PhaseModulate}
\psi = \frac{2 \pi L_\text{PS} \Delta n_{\text{eff}}}{\lambda}
\end{equation}
where $\mathit{\lambda}$ and $L_\text{PS}$ represent operating wavelength and effective modulation length. Within the C-band and O-band commonly used in silicon photonics, the transmission characteristic demonstrates a low sensitivity to $\mathit{\lambda}$ over a range of several tens of nanometers. Particularly for thermo-optic devices, the thermal modulation coefficient d$n_{\text{eff}}$/dT shows a slight increase corresponding to the increase in wavelength, which compensates for the decrease in wave number $2\pi/\lambda$~\cite{dwivedi2015experimental}. Therefore, an MZM followed by a phase shifter can achieve parallel modulation of multiple signals with different wavelengths based on WDM. However, it should be noted that waveguides induce different phase changes at various $\mathit{\lambda}$, which must be considered when using WDM in coherent ONNs.

Thanks to the phase encoding mechanism, coherent ONNs can perform multiplication with full-range input operands. Besides the SVD approach, a more intuitive example is that the sign and absolute value of the scalar can be encoded by the phase shifter and MZM separately. Following this methodology, Ref~\cite{mourgias2022noise} presents a photonic neuron architecture enabling full-range dot product operations. However, this architecture also faces limitations in terms of complexity and scalability. Specifically, implementing SVD required a matrix pre-processing step for phase mapping, which consumes extra time and power. In addition, the number of MZIs escalates quadratically with matrix size, leading to scalability issues due to the larger footprint and accumulated losses. Moreover, the phase control of the coherence network presents a challenge as well. Within the MZI mesh, each $U(2)$ operation demands a minimum of two active phase modulators. Without a reference signal, phase calibration becomes more complex than amplitude calibration (which can be straightforwardly measured using photodetectors), especially considering the fabrication deviation across individual devices and waveguides. However, some studies demonstrate the feasibility of addressing the phase calibration issue through on-chip training~\cite{zhang2021efficient,cem2023data}.

Incoherent PICs offer an alternative approach to execute tensor operations. In the absence of coherence requirements, an incoherent ONN enables broader flexibility in employing multiplexing techniques. Because of its wavelength-dependent transmission characteristics, MRRs are widely employed for encoding weights and inputs in WDM-based incoherent ONNs. From the perspective of physical design, the small footprint of MRR (achieving a radius \textless 10 µm) affords a compact layout and an enhanced scalability. As shown in Fig. \ref{SynapseImplement}.c, the add-drop MRR is a four-port optical device that consists of a microring evanescently coupled to two bus waveguides. The transfer characteristics at the through port and drop port can be expressed as Eq. (\ref{MRRTransT}) and Eq. (\ref{MRRTransD}).
\begin{equation}\label{MRRTransT}
T_t = \left| \frac{E_t}{E_{in}} \right|^2 = \frac{t_1^2 + t_2^2 \alpha_{rt}^2 - 2t_1 t_2 \alpha_{rt} \cos{\phi_{rt}}}{1 + t_1^2 t_2^2 \alpha_{rt}^2 - 2t_1 t_2 \alpha_{rt} \cos{\phi_{rt}}}\\
\end{equation}
\begin{equation}\label{MRRTransD}
T_d = \left| \frac{E_d}{E_{in}} \right|^2 = \frac{\kappa_1^2 \kappa_2^2 \alpha_{rt}}{1 + t_1^2 t_2^2 \alpha_{rt}^2 - 2t_1 t_2 \alpha_{rt} \cos{\phi_{rt}}}
\end{equation}

Here, $\kappa$ and $t$ represent the field coupling factor and transmission factor, respectively. $\alpha_{rt} = exp\left( -\alpha \cdot 2\pi R \right)$ and $\phi_{rt} = \beta \cdot 2\pi R$ are the round-trip field attenuation factor and phase, with $\alpha$ and $\beta$ being the real and imaginary parts of the complex transmission coefficients~\cite{van2016optical}. The resonance condition is $\phi_{rt} = 2m\pi$, i.e., $\lambda_{\text{res}} = 2\pi n_{\text{eff}}R/m$, where $m$ is a positive integer. A potential issue in achieving full-range modulation (i.e., from 0 to 1) arises, as it can only be attained with symmetric, lossless coupling ($\alpha_{rt} = 1$, $t_1 = t_2$), which is unrealistic for real devices. An all-pass MRR can be regarded as a special case of an add-drop MRR, distinguished by the absence of a drop bus waveguide with $t_2 = 1$ and $\kappa_2 = 0$ for Eq. (\ref{MRRTransT}). While the phase at the through/drop port can also be derived from Eq. (\ref{MRRTransT}) and Eq. (\ref{MRRTransD}), the phase response and tuning range are highly sensitive to coupling conditions and round-trip loss. Hence, MRRs are primarily employed for amplitude modulation in incoherent systems. Another challenge comes from the experimental perspective: MRRs, particularly those with high quality factors, are sensitive to environmental factors such as temperature variations and vibrations. For a more in-depth theoretical analysis of MRRs, readers may refer to relevant literature and books~\cite{van2016optical,bogaerts2012silicon}.

In tensor operations, MRRs can be modulated by tuning coupling coefficients or the round-trip phase~\cite{dong2010wavelength}. Unlike MZIs, MRRs demonstrate significant wavelength selectivity. An MRR with a moderate quality factor can achieve a resonance peak with a 3 dB bandwidth of a few hundred picometers. The narrow bandwidth makes MRRs particularly effective as tunable filters in WDM-based PICs, where they are often arranged in series for the selective modulation of signals with different wavelengths. This application is exemplified by the concept of “weight banks” proposed by Tait et al.~\cite{tait2017neuromorphic,tait2016microring
}, as shown in Fig. \ref{SynapseImplement}.d. In this architecture, photodetectors can spontaneously perform the sum of operands encoded on different wavelengths. Another incoherent architecture is the crossbar array, which implements MVMs using its programmable transfer matrix in the form of a tunable switch array~\cite{ohno2022si,feldmann2021parallel,ning2024hardware}. Ohno, et al. demonstrated a 4$\times$4 add-drop MRR crossbar array for MVMs (Fig. \ref{SynapseImplement}.e)~\cite{ohno2022si}. Specifically, each wavelength coming from the row bus waveguide can be weighted and subsequently directed into a specific column by the MRR array. The weighted elements can be collected and aggregated by a photodetector at the end of the column, thus carrying out MVM.

A notable constraint of incoherent ONNs is the challenge of directly implementing negative operands due to the amplitude modulation mechanism. An intuitive solution is dividing the matrix into positive and negative parts, subsequently mapping into the network up to 4 times to execute $(X_+ - X_-)(Y_+ - Y_-)$, and thereafter deducting the outcomes within the electrical domain. Alternatively, the positive and negative components can be processed simultaneously through two identical incoherent hardware networks. Both strategies, however, necessitate either extended processing time or increased hardware resources. Another strategy leverages the complementary output from the through and drop ports of the add-drop, enabling the full-range output as demonstrated in the work of Tait et al~\cite{tait2017neuromorphic,tait2016microring}. For the last two approaches, the post-processing subtraction can be physically implemented using balanced photodetectors, as illustrated in Fig \ref{SynapseImplement}.d.

\subsubsection{Static and dynamic weight encoding}
The above-mentioned architectures topically map one operand of MVMs (typically a weight matrix \textbf{\textit{W}}) onto hardware, executing multiplication through the transmission matrix of the PIC. Therefore, the operational speed of an ONN is, in theory, only determined by the modulation rate of input signals \textbf{\textit{x}}. However, practical implementations face several limitations. Primarily, deploying MVMs with large dimensions is challenging due to the cost, complexity, loss, and other scalability issues of PICs. While multiplexing techniques can enhance the parallelism of ONNs, time-domain hardware reuse is necessary to fulfill the computational requirements of complex machine learning tasks. This indicates that the “weight-static” architecture constitutes a bottleneck for large networks because of the huge gap between ultra-fast optical computing and slow mapping/reprogramming. To fully unleash the potential of optical analog computing, the critical role of dynamic encoding for specific tasks should be recognized. For dynamically-operated ONNs, two essential requirements must be satisfied. First, the parameters in the network need to be high-speed reprogrammable for efficient hardware reuse. This requires that the modulators operate at gigahertz-level rates, as exemplified by those based on field-effect tuning mechanisms. Secondly, dynamic operation mandates that parameter mapping and output reading be conducted “directly” without additional signal preprocessing or data correction. For instance, to map a 12×12 matrix to an MZI-mesh framework requires $\sim$1.5 ms to perform SVD and phase decomposition on a CPU~\cite{zhu2023dota}, which precludes dynamic operation due to the delay.

The importance of dynamic encoding is further highlighted by its compatibility with \textit{Transformer}~\cite{vaswani2017attention}. Transformer and the unique attention mechanisms have attracted significant research interest in recent years due to their exceptional performance in natural language processing (NLP), machine vision (MV), and large-scale language models (LLM). Unlike weight-static architectures, Transformer employs the multi-head self-attention (MHA) mechanism within its encoder and decoder blocks, necessitating matrix multiplication involving two dynamic, full-range operands. Zhu et al. presented a dynamically-operated PTC for the first photonic Transformer accelerator, leveraging coherent light interference and WDM~\cite{zhu2023dota}. As shown in Fig. \ref{SynapseImplement}.f, the elements within the input vectors \textbf{\textit{x}} and \textbf{\textit{y}} are encoded at distinct wavelengths, and then add a $-90^\circ$
phase shift to one vector. Through a directional coupler, two orthogonal signals can be recombined into the form (\textbf{\textit{x}} $\pm$ \textbf{\textit{y}}), facilitating the computation of the dot product via a balanced photodetector.

\subsubsection{Hardware-efficient PTCs}\label{HardwareEfficientPTCs}
\label{subsubsec:efficent_ptc}
For the aforementioned structures, the total number of optical components required to implement an $m \times n$ general matrix is similar, e.g., $m(m-1)/2 + n(n-1)/2 + \max(m, n)$ MZIs for an SVD-based MZI mesh, and $m \times n$ MRRs in microring-based ONNs. In addition to the high hardware costs and inherent control complexities, the required electrical components make up a large proportion of energy consumption, particularly for high-speed and high-resolution modulation. To address these challenges, researchers are exploring strategies to enhance hardware efficiency across multiple levels—from the device to the architecture—to improve the scalability of ONNs.

\begin{figure*}[htb]
  \centering
  \vspace{-5pt}
  \includegraphics[width=1\textwidth]{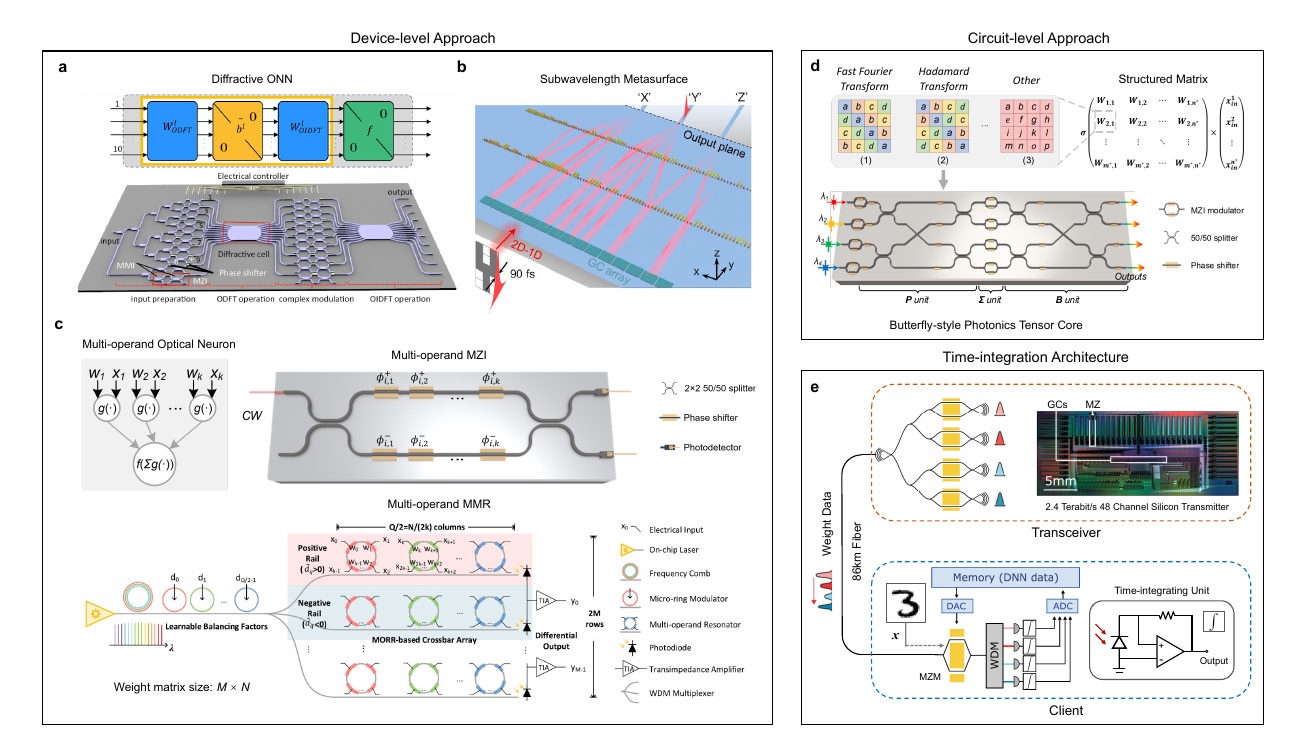}
  \vspace{-5pt}
  \caption{Implementation of hardware-efficient PTCs. (a)\&(b) Schematics of space-efficient ONNs using compact diffractive cell~\cite{zhu2022space} and metasurface structure~\cite{wang2022integrated}, respectively. O(I)DFT: Optical (inverse) discrete Fourier transform (c) Diagram of a multi-operand optical neuron, along with schematics of muti-operand MZI~\cite{feng2024integrated} and muti-operand MRR~\cite{gu2022squeezelight}. (d) The general architecture of the subspace ONNs (top), and schematic of 4$\times$4 butterfly-style tensor core (bottom)~\cite{feng2022compact}. Here, \textbf{\textit{B}} and \textbf{\textit{P}} are both unitary matrices and $\mathit{\mathbf{\Sigma}}$ is a diagonal matrix. Here, \textbf{\textit{B}} and \textbf{\textit{P}} represent butterfly-style transform unit and projection unit, respectively (e) Architecture diagram of a delocalized time-integration optical computing system consists of smart transceivers as cloud infrastructure and the edge client devices\cite{sludds2022delocalized}. The parameters of the neural network model are encoded using WDM on the cloud side and transmitted to client nodes via a long-haul optical fiber. The MVMs are carried out in time-integrating of photocurrent generated by the photodetector in the client device.}
  \vspace{-10pt}
  \label{HardwareEfficient}
\end{figure*}

Firstly, optical devices or structures featuring compact topology have been proposed to reduce PIC footprint and improve hardware efficiency. Zhu et al. developed an ONN architecture consisting of two diffractive cells and MZIs exhibiting linear scaling in relation to input dimensions~\cite{zhu2022space}, as shown in Fig. \ref{HardwareEfficient}.a. The star-shaped diffractive cell shows the capability of performing on-chip parallel Fourier transform and convolution operations. In a similar vein, an ONN leveraging the combination of WDM and MMIs has been developed and demonstrates an accuracy of 92.17\% on the MNIST dataset. Beyond waveguide-based PIC, Wang et al. proposed a metasurface-based processing unit that provides ultra-high throughput for MVM~\cite{wang2022integrated} (Fig. \ref{HardwareEfficient}.b). Within the subwavelength structure, each pair of slots acts as a weight element and connects to the following layers via in-plane diffraction and interference. For a specific neural network, the width and length of the slot need to be designed to map the corresponding weights.

The passive ONNs not only yield smaller footprints and enhanced hardware efficiency but also lower power consumption. However, a significant challenge associated with the passive task-specific PTCs is their fixed configuration, which can lead to degraded effectiveness when applied to varying tasks. To address this challenge, active multi-operand devices present another opportunity to break the fundamental limitation to achieve high-density tensor operation by squeezing MAC operation into a single device. Specifically, the execution of the length-\textit{k} dot product between input vector $\bm{x}_{\text{in}}$ and weight vector $\bm{w}$ by \textit{k}-operand devices can be represented as follows:
\begin{equation}\label{MOON_Eq}
x_{\text{out}} = f(\bm{w} \cdot \bm{x}_{\text{in}}) = f\left(\sum_{i=1}^{k} g(w_i, x_i)\right)
\end{equation}
where the function $f(\cdot)$ represents the E-O transmission function, while $g(w, x)$ is related to the encoding mechanism. Here, $\bm{w}$ can be encoded using programmable resistances (such as memristors or PCMs), tunable amplifiers/attenuators, or the effective modulation length in cases, while $\bm{x}_{\text{in}}$ can be encoded by electrical signals. The type of multi-operand devices can either be the modulators mentioned above or based on other passive optical components such as MMI~\cite{gu2023m3icro}. Feng et al. first demonstrated this method experimentally on a 4-operand MZI, achieving a measured accuracy of 85.9\% in SVHN recognition tasks with 4-bit control precision (Fig. \ref{HardwareEfficient}.c). Gu et al. proposed a compact ONN architecture based on multi-operand MRRs (MOMRR)~\cite{feng2024integrated}. The MOMRR-based ONN supports weight encoding through two sets of rails, with full-range results carried out by balanced photodetectors~\cite{gu2022squeezelight}. Additionally, the architecture can be combined with the structured pruning strategy to further improve network scalability. Theoretically, multi-operand optical synapses could execute vector operations with nearly the same footprint as the single-operand counterparts. However, the intrinsic nonlinearity and possible crosstalk among operands could present challenges for training and calibration. 

Beyond improvements at the device level, another promising approach enhances hardware efficiency at the circuit level by software-hardware co-design approaches. Subspace neural networks, for instance, sacrifice a portion of matrix representability in exchange for fewer parameters instead of implementing universal linear operations or general matrix multiplication (GEMM). The effectiveness of this strategy can extend to ONNs by trading the universality of weight representation for higher hardware efficiency. An example is the butterfly-style PTC~\cite{feng2022compact} (Fig. \ref{HardwareEfficient}.d). By parameter pruning, this architecture could be implemented using significantly fewer MZIs with a scale factor of $O(nlog_2^n)$ rather than the $O(n^2)$ required by an MZI mesh for GEMM. This study experimentally showcases a measured accuracy of over 94\% on the MNIST hand-written digits classification task, with up to 7$\times$ fewer active optical components, a 3.3× smaller footprint, and a 5.5$\times$ lower latency compared to conventional MZI mesh. A similar strategy, utilizing an MRR-based crossbar array to implement block-circulant matrices, has been demonstrated in Ref \cite{ning2024hardware}. For subspace networks, the trade-off between hardware efficiency and matrix representability is an important consideration, and more details will be discussed in Section~\ref{sec:area}. 

Beyond the one-shot broadcast architecture, an alternative method leverages time-domain integration with fewer optical components to perform MACs~\cite{sludds2022delocalized,zhang2024tempo}. Following this approach, Sludds et al. developed a delocalized optical computing system that showcases edge computing capabilities over a span of 86 kilometers~\cite{sludds2022delocalized}. As shown in Fig. \ref{HardwareEfficient}.e, the modulated WDM signals are separated by a passive demultiplexer and then fed to a set of time-integrating receivers. On the client side, only one MZM, ADC, and DAC are used, which allows an ultra-low power assumption of femtojoules per MAC operation. The photocurrent $I(t)$ generated by the photodetector produces a voltage across the integrating capacitor $C$ by accumulating charge, thereby achieving the summation operation, which can be expressed as Eq.~\eqref{TimeInteg}.
\begin{equation}\label{TimeInteg}
V_{\text{out}} = \int \frac{I(t)}{C} \, dt \propto \sum w_i x_i
\end{equation}
This strategy sacrifices the speed benefit of optical computing to achieve a smaller chip footprint and reduced hardware complexity, which enables the use of milliwatt-class edge devices.
\subsection{Implementations of Nonlinearities}
Activation functions introduce the necessary nonlinearity into the network. For PIC-based photonic neurons, the implementations of nonlinearities fall into two major categories: the optical-electrical-optical (O-E-O) and the all-optical approach.
\subsubsection{O-E-O nonlinearities}
The activation function within O-E-O neurons can be realized most directly by routing the weighted sum through A/D conversion into digital processing units, such as CPUs, GPUs, or FPGAs. This process performs nonlinear operations digitally, and then converts the digital output back into analog signals that are subsequently fed into photonic neurons. The primary advantage of this approach is its extensive transfer ability of existing ANN architectures to the PIC platform, which also facilitates the implementation of arbitrary activation functions. However, the compulsory A/D and D/A conversion, along with digital processing, impose latency and power consumption, thereby undermining the efficiency advantage inherent in ONNs. As mentioned in Section \ref{section2}, the performance and speed of an optical computing system are constrained by its weakest link, which in this case is the E-O interface and the digital processing. Alternatively, nonlinearities can also occur in the analog domain by leveraging the inherent nonlinear responses of specific electronic components or analog circuits~\cite{chen2023full,vatalaro2021low}.

Beyond implementing nonlinearity purely in the electronic domain, nonlinearity can also be introduced within the E-O/O-E conversion processes, such as the nonlinear transmission properties of E-O modulators during encoding. For instance, the nonlinear part of the cosine term in the MZM transfer function exhibits similarities to the sigmoid function. This methodology has been validated on the last two layers of an ONN for the MNIST classification task in Ref~\cite{mourgias2022noise}, and a similar approach involving the built-in nonlinearity of MRRs has been proposed by Gu, et al.~\cite{gu2022squeezelight}.
The main advantage of introducing nonlinearity in encoding is it does not need extra O-E conversions. However, since the nonlinearity is entirely contingent on the transmission characteristics of E-O modulators, this dependency may pose challenges during the training process. Inappropriate activation functions, particularly in deep networks, can result in gradient vanishing or explosion and low training efficiency.
To achieve reconfigurable activation functions, Williamson et al. introduced a nonlinear unit that converts a small portion of the optical output into an electrical signal to modulate the original optical signal (Fig. \ref{Nonlinearities}.a)~\cite{williamson2019reprogrammable,fard2020experimental}. This setup offers two approaches for the electrical part. The first approach converts the tapped signal into an electrical signal directly, allowing for moderate adjustments in nonlinearity through varying electrical biases. The second strategy utilizes reconfigurable lookup tables controlled by a microcontroller, which enables the generation of arbitrary nonlinearities and aligns more closely with an all-electrical approach. Likewise, MRRs can be employed to realize this methodology. In Ref~\cite{ashtiani2022onchip} and~\cite{tait2019silicon}, activation functions are implemented by modulating a CW probe with the output from photodetectors (Fig. \ref{Nonlinearities}.b). While this architecture avoids the loss accumulation issue, it requires an additional laser source as the probe. In addition to introducing nonlinearity from E-O modulation, O-E conversion processes, exemplified by the response of photodetectors, can also serve as the source of nonlinearity~\cite{shi2022nonlinear}. Nevertheless, these approaches are also constrained by limited reconfigurability.

\subsubsection{All-optical nonlinearities}
\label{AllOptical}
\begin{figure*}[htb]
  \centering
  \vspace{-5pt}
  \includegraphics[width=0.95\textwidth]{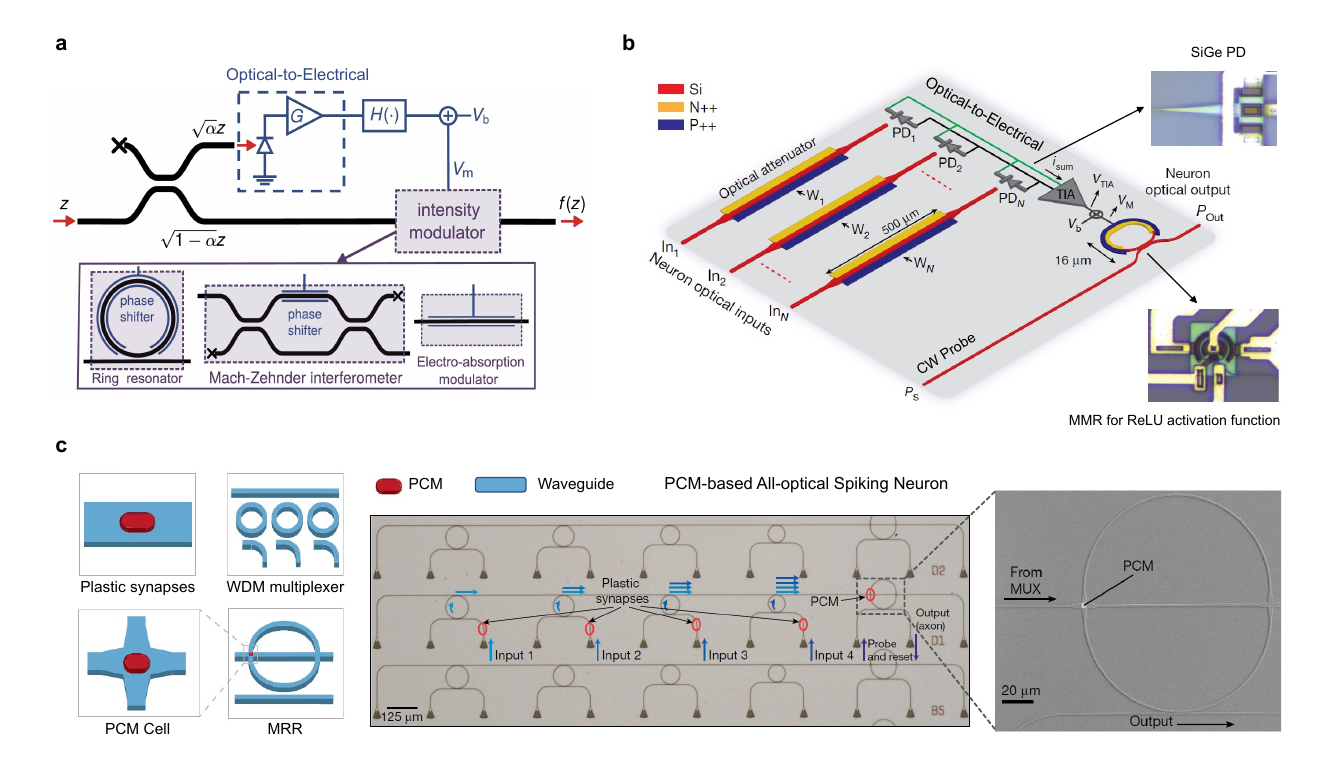}
  \vspace{-5pt}
  \caption{On-chip implementations of programmable nonlinear activation functions. (a) \& (b) Schematics of O-E-O nonlinear units for realization of reconfigurable activation functions~\cite{williamson2019reprogrammable,fard2020experimental,ashtiani2022onchip}. (c) A PCM-based all-optical spiking neuron~\cite{feldmann2019all}. The input spikes are weighted via PCM cells, and the PCM cell on the MRR could switch the resonate condition when the accumulated power of the postsynaptic spikes exceeds a predefined threshold, thereby controlling the generation of output spikes.}
  \vspace{-10pt}
  \label{Nonlinearities}
\end{figure*}

Without electrical components, all-optical approaches implement activation functions via the nonlinear response of materials or devices to optical signals. Several ONNs have been demonstrated using semiconductor optical amplifiers (SOAs)~\cite{kravtsov2011ultrafast}, saturable absorbers~\cite{nahmias2013leaky,shi2023photonic}, and techniques exhibiting excitable behavior~\cite{romeira2013excitability}. All-optical nonlinearities can also be achieved by PCMs, such as the all-optical neuron developed by Feldmann et al~\cite{feldmann2019all}. In this work, the PCM, functioning as a waveguide cladding, is used to modulate the pre-synaptic input and govern the resonance state of MRR, thereby controlling the generation of output spikes (Fig. \ref{Nonlinearities}.c). Beyond the issues of activation function applicability previously discussed, the implementation of all-optical neurons faces several other challenges. Firstly, even though extra electrical devices are not required, the substantial power required to excite nonlinearities does not offer any power consumption advantages compared to the O-E-O approaches. Additionally, the fabrication and integration of optical components for all-optical nonlinearity, such as PCM, excitable lasers, and optical amplifiers, pose challenges as well. Lastly, since both the control signal and post-synaptic output are optical signals, a potential issue exists in distinguishing the response from the control signals or bias.

\subsection{Optical Neural Network Training}
\label{subsec:onn_training}
As an analog computing platform, PICs for AI applications are inherently susceptible to various non-ideal conditions, such as environmental changes, process variation, and limited control precision. 
These factors can decrease inference accuracy and potentially degrade the signal-to-noise ratio (SNR) of computations.
To mitigate the decrease in accuracy, it is crucial to train ONNs with careful consideration of non-idealities that may occur during inference, making it a ``circuit-aware'' approach will enhance its noise resilience.
The current ONN training can be classified into two categories: hardware-aware \emph{ex-situ} training, which involves training with the help of digital computers, and on-chip \emph{in-situ} training.
\subsubsection{Hardware-aware ex-situ Training}
\begin{figure}[htb]
    \centering
    \vspace{-5pt}
\includegraphics[width=1\columnwidth]{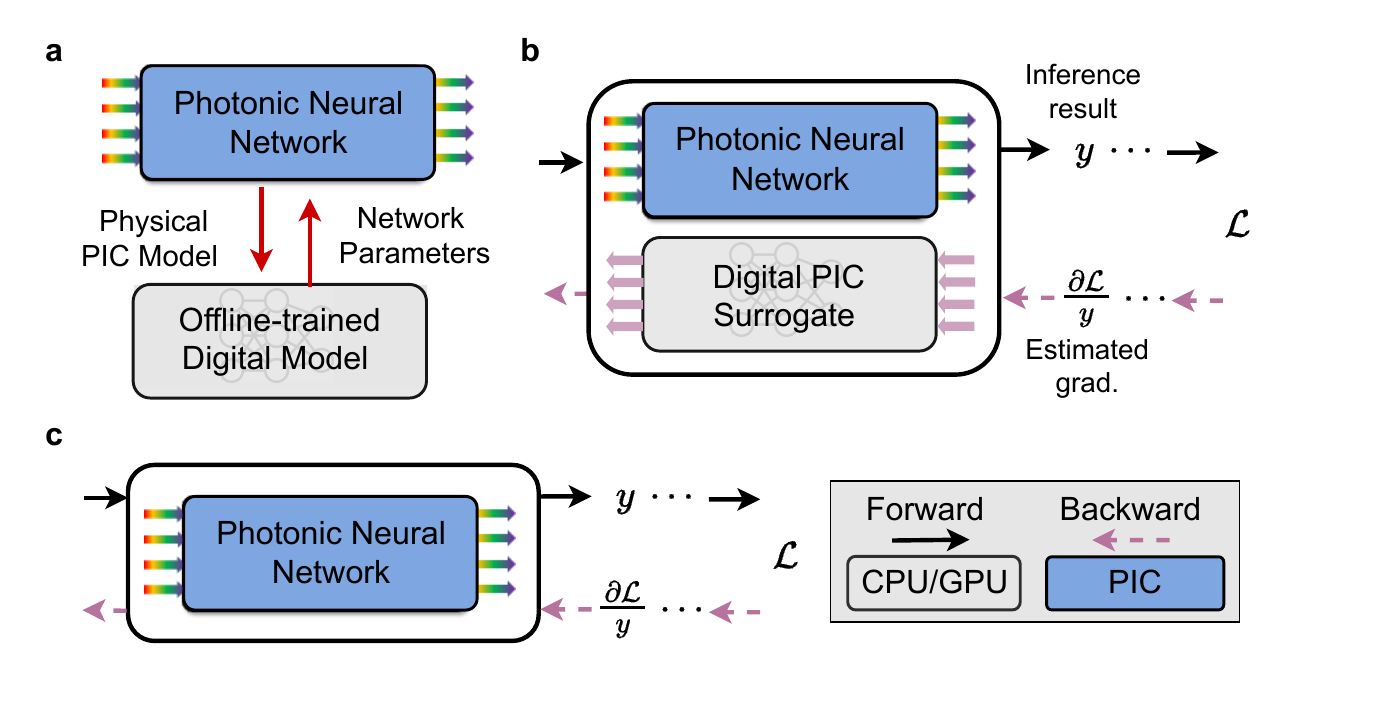}
    \vspace{-5pt}
    \caption{Optical neural network training methods. (a) Offline training on a digital platform, such as a GPU, mimics the behavior of analog hardware by building a physical model of PICs. (b) Physical neural network training, which utilizes PICs for forward pass and incorporates a differentiable digital model for backpropagation.
    (c) \emph{In-situ} optical neural network training is performed entirely on photonic hardware.
    }
    \label{fig:compare_training}
    \vspace{-10pt}
\end{figure}

Hardware-aware \emph{ex-situ} training offloads the training process to digital computers and utilizes various hardware-aware training techniques to capture PIC behavior as precisely as possible during training.
One commonly used technique is noise-aware training~\cite{zhao2019design,gu2020roq}, as shown in Fig.~\ref{fig:compare_training}.a. 
This approach involves modeling the behavior of PICs while considering various non-ideal effects. 
Subsequently, the PIC model is injected into the training to reduce the gap between training and real inference.
Gaussian noise is commonly used to model various noise sources in photonic systems. 
For instance, shot noise and thermal noise are modeled using a Gaussian distribution by measuring on-chip photonic multiplication results and fitting their distribution~\cite{sludds2022delocalized}.
In addition to injecting non-idealities into the training, some work explicitly models the transfer matrix of photonic neurons and embeds them in the forward computation pass during training~\cite{feng2024integrated,gu2022squeezelight}.
This injection is crucial because it introduces a unique nonlinearity term that ONNs need to be aware of in order to accurately capture the behavior of photonic neurons.
Moreover, in Ref~\cite{zhu2023dota}, the noisy transfer matrix is derived and explicitly injected in the training under some noise assumptions.
However, this approach encounters two main challenges. Firstly, accurately representing all on-chip non-idealities poses a significant difficulty, and environmental
fluctuations can further affect inference accuracy. 
Second, the computational overhead required to model the physical system accurately can be exceedingly high or even prohibitive, making the training very slow and costly.

Aware of these challenges, some works advocate for training optical neural networks directly with non-ideal physical responses, called physical network training, as shown in Fig.~\ref{fig:compare_training}.b. 
The forward pass is executed on PICs and the loss signal $\mathcal{L}$ is obtained by comparing the physical and intended outputs.
However, a challenge arises during the backward pass due to the undifferentiable nature of the physical response.
To address this challenge, the idea of adopting a differentiable PIC surrogate model in digital domain has been proposed~\cite{feng2022compact, wright2022deep, zhan2024physics}.
With the differentiable model, the gradient of the loss can be propagated back with respect to the controllable parameters.
This strategy obviates the need for tedious modeling and analysis of on-chip noise sources, incorporating the noisy behavior of photonic chips naturally during training.

\subsubsection{In-situ Training}

\emph{In-situ} training aims to perform training directly on-chip, enabling the inherent consideration of all kinds of on-chip non-idealities, as shown in Fig.~\ref{fig:compare_training}.c.
This approach can potentially boost accuracy to the greatest extent by directly incorporating the real-world behavior of photonic hardware into the training process.

Although \emph{in-situ} training is \textit{straightforward} and \textit{ideal}, it is challenging to implement directly on the optical computing platform.
Considering an optical neural network layer $l$, optical components execute the linear projection $W^{l}(\mathbf{\Phi}^{l})$ complemented by a digital or analog nonlinear transformation $f^{l}$. 
With the input $\mathbf{x}^{l}$, the forward manner can be defined as,
\begin{equation}
\begin{aligned}
    \mathbf{y}^{l} &= W^{l}(\mathbf{\Phi}^{l}) \mathbf{x}^{l} \\
    \mathbf{x}^{l+1} &= f^{l} (\mathbf{y}^{l})
\end{aligned}.
\end{equation}
Here, $\mathbf{\Phi}^{l}$ represents the device configurations, which are the device control variables.
For the backward pass, assuming we can access the gradient of loss over $\mathbf{x}^{l+1}$, ${\partial \mathcal{L}}/{\partial \mathbf{x}^{l+1}}$, we need to obtain the gradient over input $\mathbf{x}^{l}$ and device configurations $\mathbf{\Phi}^{l}$.
The first step is to obtain the gradient over $\mathbf{y}^{l}$, which could be prohibitive if the activation function is undifferentiable without analytical formulations, especially for customized analog activations.
After determining the gradient ${\partial \mathcal{L}}/{\partial \mathbf{y}^{l}}$, the gradients over inputs and device configurations are defined as $W^{l}(\mathbf{\Phi}^{l})^{T}  \frac{\partial \mathcal{L}}{\partial \mathbf{y}^{l}}$ and $\frac{\partial \mathcal{L}}{\partial \mathbf{y}^{l}} \mathbf{x}^{T} \frac{\partial \mathbf{W}^{l}(\mathbf{\Phi}^{l})}{\mathbf{\Phi}^{l}}$. 
The challenges of obtaining the above two items stem from several aspects.
First, it requires bidirectional input support to access the transpose of $W^{l}(\mathbf{\Phi}^{l})$.
Second, one may argue that another optical processor can be used to implement matrix multiplication in the backward pass, thereby avoiding the need for bidirectional support.
However, noise and precision limitations cause the gradient computation to deviate from the desired matrix multiplication, thus impeding convergence.
Third, obtaining the analytical gradient over the real device configurations, ${\partial \mathbf{W}^{l}(\mathbf{\Phi}^{l})}/{\mathbf{\Phi}^{l}}$, can be very complex and prohibitive, as demonstrated in the case of an MZI-ONN~\cite{gu2021l2ight}.

\begin{figure}[htb]
    \centering
    \vspace{-5pt}
\includegraphics[width=0.48\textwidth]{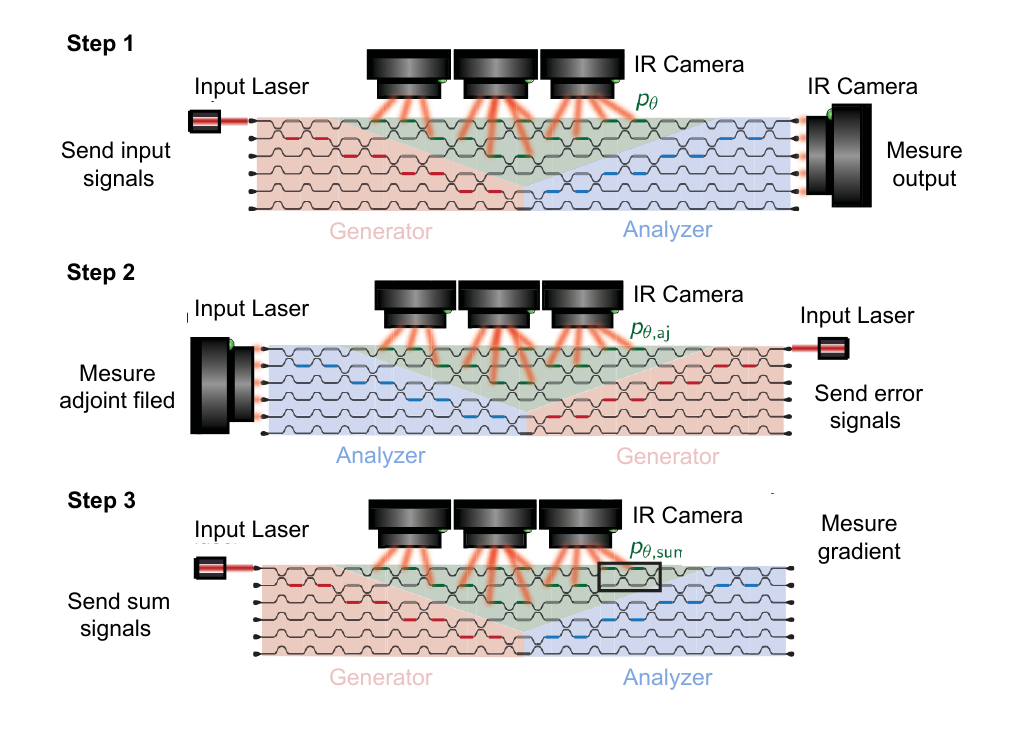}
    \vspace{-5pt}
    \caption{In-situ backpropagation concept of adjoint-variable method~\cite{pai2023experimentally}. The forward, backward, and sum steps of their backpropagation concept are shown to derive the gradient with respect to phase parameters $\theta$.
    }
    \label{fig:adjoint}
    \vspace{-10pt}
\end{figure}

An adjoint-variable method was proposed theoretically to implement on-chip backpropagation by interfering with adjoint and forward fields~\cite{hughes2018training}. 
Recent work has implemented the concept of in-situ backpropagation within a triangular MZI mesh~\cite{pai2023experimentally}.
As depicted in Fig.~\ref{fig:adjoint}, the method involves forward and backward signal propagation, followed by gradient calculation.
In step 1 and step 2, the “forward inference” signal $x$ and “backward adjoint” signal $x_{aj}$ are sent forward, respectively.
Then the “sum” vector $x-i (x_{aj})^{*}$ is send forward again.
The gradient with respect to the phase variables can be finally obtained.
However, this implementation requires additional power/phase monitors, faster microcontrollers, and more precise detectors, which increases hardware control complexity and imposes scalability concerns.
Additionally, Ohno et al. attempted to design an MRR crossbar array capable of implementing matrix multiplication between the gradient and the transpose of the weight matrix~\cite{ohno2022si}, which is a fundamental operation in conventional backpropagation algorithms. While promising, this method has yet to be demonstrated on-chip.
Besides, there is a line of works on on-chip learning protocols evaluated in simulation, which will be discussed in Section~\ref{subsec:learnability}.
\section{Photonic-Electronic AI Accelerator: a Glance at the Architecture Level}\label{section4}
Photonic AI computing is experiencing rapid advancements in both device and circuit-level innovations. 
To fully harness the potential of optical computing, it's imperative to develop comprehensive systems that combine PIC-based computing with key auxiliary components, such as memory and datapath elements. 
This necessitates architecture-level efforts to translate circuit-level innovations into practical frameworks suitable for real-world applications.
However, the architecture-level study of photonic AI accelerators is still in its infancy, and limited research has been conducted in this direction.
Given the significance of comprehending photonic AI accelerators, this section provides an overview from an architectural perspective, focusing deeply on the system components and covering various design considerations.


\subsection{Fully-optical vs. Photonic-electronic Accelerators}
\subsubsection{Fully-optical accelerator}
A fully optical accelerator refers to implementing all operations, including both computation and necessary nonlinear activation functions, entirely within the optical domain.
Recent works have demonstrated the integration of optical computation and on-chip nonlinearity~\cite{mourgias2019all,crnjanski2021adaptive,feldmann2019all,xiang2020all}.
Fully optical accelerators, although promising high bandwidth without the power consumption of E-O interference, still confront significant challenges due to scalability and practical implementation concerns.
Firstly, the on-chip nonlinearity is not yet a mature scheme with low energy efficiency or flexibility compared to electronics, as discussed in Section~\ref{AllOptical}.
Secondly, the significant loss imposes a substantial requirement for optical power to meet the detection threshold.
Additionally, computation errors will accumulate along the optical computing layers, deviating the final outcome significantly from the expected value.

\subsubsection{A more practical paradigm: photonic-electronic accelerator}
Given the challenges associated with practically implementing all-optical accelerators, the photonic-electronic hybrid accelerator emerges as a more feasible and competitive photonic AI solution~\cite{zhou2022photonic}, which is also the key focus in this section.
The current hybrid accelerator setup takes advantage of both emerging photonics and mature electronics and builds a system with a tight integration of photonic and electronic integrated circuits.
The intensive tensor operations are executed on optical parts in the analog domain, while the electronic segment includes digital memory for data storage and distribution, as well as essential units for data writing/reading, flow control, and minor data processing.
Combining digital and analog domains results in a mixed-signal setup, therefore, requiring E-O/O-E, D/A, and A/D conversions.
Although the conversion processes incur additional power consumption and delay, they also offer some advantages.
For example, the A/D conversion process can be viewed as a denoising step, as it involves discretizing the continuous analog signals into digital representations.
This discretization helps filter out noise that is present in the analog signals, avoiding error propagation.

\subsection{Architecture and Workload Mapping}
\subsubsection{Architecture}
\begin{figure}[htbp]
  \centering
  \vspace{-5pt}
  \includegraphics[width=0.45\textwidth]{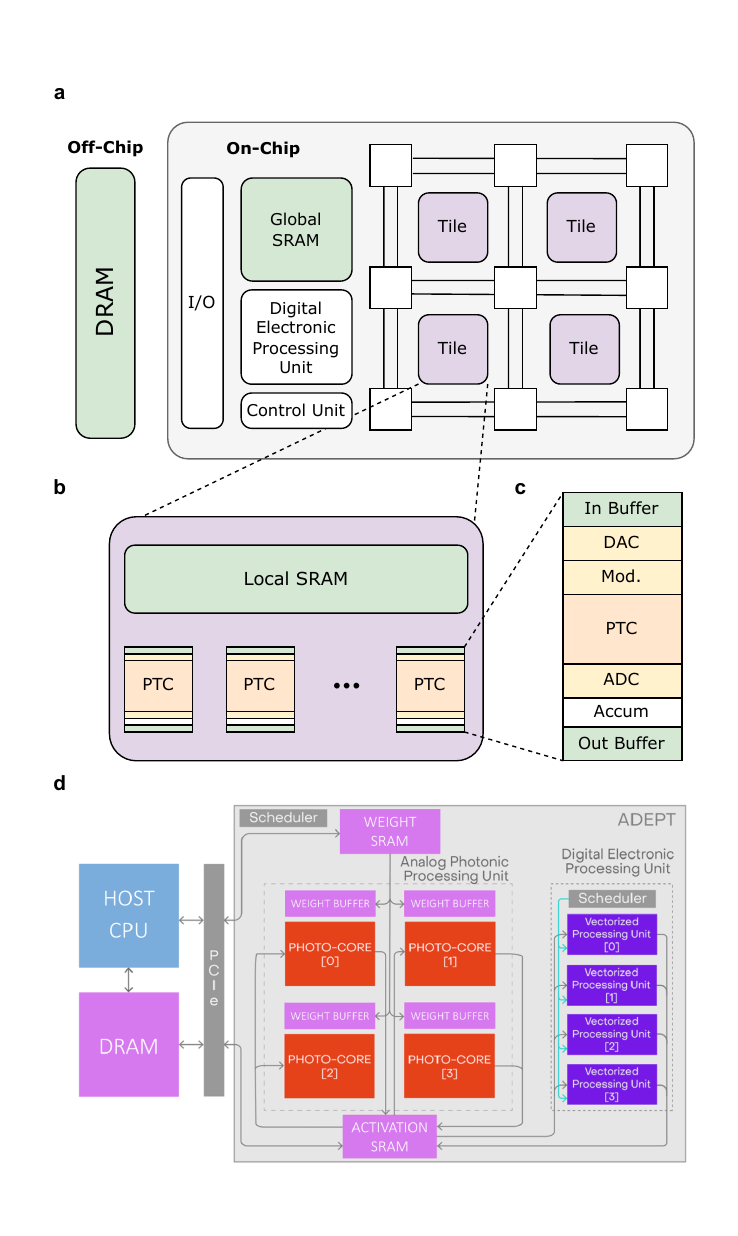}
  \vspace{-10pt}
  \caption{The generic photonic accelerator architecture. (a) A high-level representation of the full system architecture, including off-chip and on-chip components. (b) The architecture representation of a single tile. (c) The details of the linear computing unit using photonic components. (d) The system architecture of ADEPT interacting with peripheral components~\cite{demirkiran2023electro}.}
  \vspace{-10pt}
  \label{ONN_arch}
\end{figure}

The photonic-electronic accelerator can be classified into two types based on its application, which range from task-specific accelerators, such as those for CNNs~\cite{shiflett2020pixel, shiflett2021albireo,li2023photofourier}, to general-purpose architectures~\cite{demirkiran2023electro, zhu2023dota}.
Despite their varied applications, these accelerators exhibit a similar generic high-level micro-architecture, as shown in Fig.~\ref{ONN_arch}. The figure describes the generic architecture in a top-down manner, from the top view of an on-chip accelerator to interaction with peripheral devices, such as off-chip dynamic random-access memory (DRAM). Within the accelerator, the chip node contains both the photonic and electronic circuits.
The photonic part handles tensor operations, while the electrical component is responsible for flow management. Each tile is constructed with multiple PTCs and a shared local static random-access memory (SRAM) to store data that can be accessed by different PTCs.
The PTC necessitates several digital units to support its operation, such as the in-buffer and out-buffer for data storage, ADC, DAC, and the modulator for signal conversion.
\noindent\textbf{}

\noindent\textbf{Memory}\quad
Memory design is a critical aspect of modern accelerator design, given that data movement frequently constitutes the bottleneck of the entire system. This challenge is observed not only in traditional electronic processors but also in optical processors.~\cite{ramey2020silicon}. In addition to addressing costly data movement, the ultra-fast characteristic of photonic computing requires specialized memory design to accommodate the requirements for high-speed data access.

Several key design considerations can be taken into account during the memory system design process:
(1) Memory hierarchy: Adopting a memory system with multiple hierarchical levels is a common strategy to reduce memory access costs and meet the speed requirements. Previous studies have organized memory systems from external off-chip DRAM to internal on-chip SRAM~\cite{demirkiran2023electro, lam2024dynamic, zhu2023dota, shiflett2020pixel, shiflett2021albireo}. This progression typically involves using global SRAM to drive multiple local SRAMs that store frequently accessed data.
For example, Lightmatter has introduced an architecture denominated ADEPT as shown in Fig.~\ref{ONN_arch}.d~\cite{demirkiran2023electro}, featuring separate SRAMs for activations and weights to accommodate their differing access frequencies and patterns. 
Specifically, weight buffers are utilized to facilitate communication between the large but slow global SRAMs and the ultra-fast PTCs, enabling parallel programming of weight blocks.
(2) Bandwidth matching: Towards efficient data transfer, bandwidth matching is critical to ensure that the memory bandwidth aligns with the requirements of PTCs.
One effective approach is to integrate multiple memory channels, as demonstrated in Ref~\cite{demirkiran2023electro, zhu2023dota}.
To meet latency demands and enable efficient transmission of parameter blocks, the large global SRAM block is partitioned into smaller sub-arrays. This allows data to be read at brief intervals, aligning actual data access latency with specified requirements.
(3) Data prefetching: Given the regularity of neural network workloads, future memory access can be anticipated, allowing for the use of double buffering to hide data loading latency, as demonstrated in  Ref~\cite{zhu2023dota}.
(4) Data reuse: A common approach in photonic accelerators, known as 'weight static', explores weight locality to enhance reuse and minimize data movement.

\noindent\textbf{Optical computation}\quad
The fundamental computing primitive is a single PTC designed to accelerate tensor operations.
The accelerator could integrate multiple PTCs, allowing for task distribution across multiple cores, thus reducing execution time and increasing throughput.
These multiple on-chip PTCs can be further organized into tiles, following a modular design principle where each tile shares associated memory and control logic. 
Different tiles communicate with each other through router logic, using network-on-chip or optical interconnect, as demonstrated in Fig.~\ref{ONN_arch}.a.

\noindent\textbf{Digital electronic processing unit}\quad
In neural network acceleration, various non-GEMM operations are essential, such as element-wise non-linear operations (e.g., ReLU, GELU, and sigmoid), reduction operations (e.g., softmax and max-pool), batch and layer normalization, and element-wise operations (e.g., bias). 
Digital electronic processing units are employed for these tasks due to their higher efficiency.
However, to retain the performance benefits of photonics, it is crucial to match the high throughput of photonic part, especially considering the challenges faced by electronic processing units operating beyond 2 GHz.
Demirkiran et al. proposed to equip each PTC with vectorized processing units and duplicate logic units internally to enhance processing speed~\cite{demirkiran2023electro}.

\subsubsection{Workload Mapping}
\label{WorkloadMapping}
This subsection provides an overview of how photonic accelerators perform neural network inference, especially when handling matrix-multiplication-related workloads.
The workload needs to be partitioned and scheduled appropriately to map the matrix multiplication to PICs.

The first mapping method is specific to convolutional layers, where the mapping honors the sliding window property of convolution. The sliding of the input feature map through the kernel can be viewed as a series of dot products between the kernel matrix and the sliding receptive fields. 
In Ref~\cite{xu2021tops}, the sliding receptive fields are sent to the photonic accelerator with a time-wavelength interleaving to enable the multiplication with kernel weights in a temporal manner. Ref~\cite{shiflett2020pixel,shiflett2021albireo} further cascade multiple sliding receptive fields with the same kernel to perform an MVM operation. However, this mapping method is specific to convolutional layers, which limits its generalizability across non-convolutional workloads.

The alternative mapping approach involves transforming the tensor multiplication task into a GEMM operation between two large matrices. This conversion is straightforward for linear layer workloads. In the case of convolutional layers, the image-to-column (img2col) technique is commonly employed to convert them into MVM tasks~\cite{chetlur2014cudnn,bagherian2018chip}.
Matrix tiling is required to partition the input matrices into smaller blocks, and the block-wise multiplication is executed on the photonic accelerator.
However, the scheduling of these small blocks, i.e., the order in which these tiles are processed in PTCs, requires careful design, which is sometimes overlooked in recent optical computing studies.
Dataflow selection has a significant impact on this scheduling. 
Currently, many optical computing platforms constrain dataflow selection to weight-stationary dataflow~\cite{shen2017deep, demirkiran2023electro}.
This is because weight programming is often slow and costly, favoring the weight-static mode.
In this scheme, block-wise weight matrices are kept in the PTC to maximize their reuse among different input data sizes. 
However, recent advancements in dynamically-operated PTC designs, such as those proposed in Ref~\cite{zhu2023dota, zhu2024coherent}, have eliminated these restrictions due to advancements in weight programmability, enabling flexible dataflow selection based on workload requirements. For example, output-stationary dataflow is employed in Ref~\cite{zhu2023dota} to support attention matrix multiplication, as the matrices have limited reuse opportunity in this dynamic matrix multiplication scenario.
\section{Toward Efficient Photonic AI Computing: A Software-hardware Co-design Perspective}
\label{section5}

In this review, we have explored the promising potential of photonics for AI acceleration.
As a highly interdisciplinary field, photonic AI computing requires contributions from various domains, including device, circuit, architecture, and algorithm levels.
Previous sections have examined a range of efforts from academia and industry across the device, circuit, and architecture parts.
However, solely focusing on progress at the individual component level is insufficient to fully unlock the vast potential of photonic computing.
Instead, a holistic approach across multiple levels is crucial for maximizing the performance of photonic AI systems, necessitating a software-hardware co-design perspective.

In this section, we will delve into the key challenges of current photonic AI systems, specifically in terms of area density, energy efficiency, noise robustness, and trainability.
We will feature representative studies that address these challenges through efforts across multiple levels of the system.
\subsection{Area}\label{sec:area}
\subsubsection{Issue}

The large spatial footprint of PICs is a significant concern, as optical devices typically have much larger physical dimensions compared to nanometer-scale transistors, spanning hundreds or thousands of square micrometers. 
In this case, PICs generally have low packing density and are not competitive in area efficiency.
Consequently, it becomes challenging to accommodate a large number of photonic components or a large matrix on a single chip, limiting hardware scalability and compute density.

\subsubsection{Co-design progress}
The concern over area cost has prompted the development of compact photonic devices and the advancement of fabrication processes. 
This, combined with various hardware-efficient PTC designs discussed in Section~\ref{HardwareEfficientPTCs}, aims to address the challenge of large spatial footprints in PICs.
Beyond the methods focusing solely on device or circuit levels, researchers are exploring holistic device-circuit-algorithm co-design efforts in two promising directions: (1) domain-specific photonic computing engines to trade off between matrix expressivity and hardware efficiency; (2) AI-assisted automatic compact PTC design.


The first line of focuses on developing domain-specific photonic computing engines~\cite{gu2020towards, feng2022compact, zhu2022space, xiao2021large, gu2023m3icro, li2023photofourier}, instead of universal linear units as in previous works, such as MZI meshes and MRR banks~\cite{shen2017deep,tait2017neuromorphic}. 
The over-parameterization of neural networks has inspired various research efforts exploring the construction of efficient neural networks beyond conventional GEMM within a restricted matrix parameter space, including low-rank neural networks and structured neural networks~\cite{oseledets2011tensor,ding2017circnn}.
We refer to such neural networks with a restricted weight matrix space as subspace neural networks~\cite{feng2022compact,ning2024hardware}.
Subspace neural networks have shown considerable improvements in efficiency while maintaining comparable representability to traditional neural networks.
The success of efficient subspace neural networks can be leveraged in ONNs by sacrificing the universality of weight representation in exchange for higher hardware efficiency.
For instance, Gu et al. proposed to implement an efficient circulant neural network and devise a novel butterfly photonic architecture with improved area efficiency over previous circuits\cite{gu2020towards, ding2017circnn}.
It implements the optical fast Fourier transform (OFFT) and its inverse (OIFFT), which are used to efficiently perform circulant matrix multiplication.
Gu et al. further extended the proposed architecture to a trainable transform structure to enable the implementation of more matrix transformation~\cite{gu2020towards}.
Moreover, to solve the issue of quadratic increase in the number of MZI devices when supporting larger matrices in an MZI mesh~\cite{shen2017deep}, Xiao et al. applied tensor-train decomposition first to decompose large over-parameterized weight matrices into smaller ones, thus substantially reducing the number of MZI devices required~\cite{xiao2021large}. 

Another noteworthy advancement in this field is the exploration of automatic PTC design, departing from the manual design paradigm. 
In this approach, the footprint can be incorporated as a constraint to enable the automatic generation of compact PTC.
For instance, ADPET introduces the first automatic AI-assisted differentiable search framework for PTC topology design~\cite{gu2022adept}. 
It first constructs a probabilistic photonic SuperMesh and then employs differentiable optimization in a huge and highly discrete PTC search space.
This framework adapts to various circuit footprint constraints and foundry PDKs. 
The PICs developed using this method demonstrate a substantial increase in footprint compactness, ranging from 2 to 30$\times$ compared to both the MZI mesh and the manually designed compact butterfly mesh. 
This automated approach promises to revolutionize the design process, enabling the creation of more efficient and compact PTCs for AI applications.
\subsection{Energy Efficiency}
\subsubsection{Issue}
Energy efficiency is a crucial metric for evaluating computing hardware. When assessing the energy efficiency of photonic AI hardware, especially given its mixed-signal setup, it is essential to consider the energy costs associated with both the digital and optical components of the system.
Following previous studies~\cite{zhu2023dota,anderson2023optical}, the energy cost of photonic computing can be broadly categorized into optical costs associated with performing computation and electrical costs related to loading operands ($X$ and $Y$) and detecting output ($O$), which can be expressed as follows:

\begin{equation}
\label{eq:energy}
    \begin{aligned}
        E&= E_{\text{laser}} +  \underbrace{E_{\text{comp}}}_{\text{ compute}} + \underbrace{E_{\text{load}}}_{\text{load X}} + \underbrace{ E_{\text{load}}}_{\text{load Y}} + \underbrace{E_{\text{det}}}_{\text{ detect O}} \\
        E_{\text{load}} &= \textbf{\textit{E}}_{\textbf{read}} + \textbf{\textit{E}}_{\textbf{DAC}} + E_{\text{mod}} \\[4pt]
        E_{\text{det}} &= E_{\text{PD}} + \textbf{\textit{E}}_{\textbf{ADC}} + E_{\text{amp}} + \textbf{\textit{E}}_{\textbf{write}} \\
    \end{aligned}.
\end{equation}
Here, $E_{\text{load}}$ encompasses the energy costs for memory reading ($E_{\text{read}}$), D/A conversion ($E_{\text{DAC}}$), and signal modulation ($E_{\text{mod}}$). $E_{\text{det}}$ represents the costs of optical signal detection ($E_{\text{PD}}$), signal amplification ($E_{\text{amp}}$), A/D conversion ($E_{\text{ADC}}$), and the subsequent writing of results back to memory ($E_{\text{write}}$). $E_{\text{comp}}$ includes other energy costs associated with performing computation, which could be negligible for a fully passive PIC. 

Among all components, the transition between digital and analog signals presents a significant bottleneck in the system energy consumption, particularly evident in data movement (memory-associated cost) and ADC/DAC, as labeled in bold in Eq. \eqref{eq:energy}. 
This transition can occupy more than 80\% of the overall energy consumption~\cite{ramey2020silicon, zhu2022fuse}, which is different from the power composition of the optical digital computing outlined in Section \ref{subsec:ReconfigurablePIC}.

\subsubsection{Co-design Progress}
Device-level advancement promises straightforward reduction in energy costs in Eq. \eqref{eq:energy}, such as the progress in energy-efficient ADCs~\cite{adc_survey}, efficient optical modulators, and on-chip laser~\cite{zhou2023prospects}.
However, in this section, we explore recent advancements beyond the device level, with a particular focus on addressing the signal conversion and memory aspects.

At the circuit and architecture levels, several efforts have been made to reduce signal conversion and data movement costs, as summarized in Table~\ref{tab:energy}. Optical broadcast is a widely-used technique that enables spatial sharing of encoded signals~\cite{tait2017neuromorphic}, thereby saving on DAC and E-O modulation costs. 
This approach has been further extended to a crossbar-style design in Ref~\cite{zhu2023dota} to enable both input and weight to be spatially shared. 
Another strategy involves keeping weights static in photonic devices or employing non-volatile devices to reuse encoded signals across different inputs temporally. Additionally, leveraging spectral parallelism allows for sharing operands among different inputs. 
Both approaches can lower the encoding costs related to DAC and modulation. 
Time-integration techniques have been employed to explore analog-domain temporal accumulation~\cite{sludds2022delocalized}, significantly reducing the A/D conversion frequency and preserving more computations within the analog domain. Recently, the use of E-O analog memory has gained attention~\cite{lam2024dynamic}, where analog memory is placed near photonic devices, and DACs are reused across rows of analog memory, thereby reducing DAC costs.

\begin{table}[t!p]
\centering
\caption{ Comparison of recent approaches on saving signal conversion cost, including D/A ($\textit{E}_{\text{DAC}}$), A/D($\textit{E}_{\text{ADC}}$), E-O($\textit{E}_{\text{mod}}$) energy costs.
}
\label{tab:energy}
\resizebox{0.93\columnwidth}{!}{
\begin{tabular}{c|c|c|c}
\hline
            Method                & Input X       & Weight Y & Output O \\
                            \hline
Optical broadcast~\cite{tait2017neuromorphic}           & $\textit{E}_{\textbf{DAC}}, \textit{E}_{\textbf{mod}}\downarrow$ & -                    & -       \\
Weight-static dataflow~\cite{demirkiran2023electro}          & -        & $\textit{E}_{\textbf{DAC}}, \textit{E}_{\textbf{mod}}\downarrow$ &         \\
Spectral parallelism~\cite{feldmann2021parallel, shiflett2021albireo}          & -        & $\textit{E}_{\textbf{DAC}}, \textit{E}_{\textbf{mod}}\downarrow$ &         \\
Time-integration~\cite{li2023photofourier, zhu2023dota, gu2023m3icro}             & -        & -                    & $\textit{E}_{\textbf{ADC}}\downarrow$     \\
Electro-Optic analog memory~\cite{lam2024dynamic}               & -        & $\textit{E}_{\textbf{DAC}}\downarrow$               &    -     \\
Non-volatile device~\cite{feldmann2021parallel, miscuglio2020photonic}&      -    & $ \textit{E}_{\textbf{mod}}\downarrow$                  &     -   \\

\hline
\end{tabular}
}
\vspace{-10pt}
\end{table}

At the algorithm level, many studies have adopted low-bit quantization techniques to preserve accuracy while using low-precision weights, inputs, and activations.~\cite{gu2020roq, gu2020towards,kirtas2023mixed, wang2022integrated}. This approach can reduce the required DAC and ADC costs with lower resolution, as well as reduce the memory data movement cost.
Furthermore, the residue number system has been explored as a method to reduce precision requirements by achieving high-precision computations using low-precision components~\cite{demirkiran2023accelerating}.

\subsection{Noise Robustness}
\subsubsection{Issue}
\begin{figure}
    \centering
    \vspace{-5pt}
\includegraphics[width=\columnwidth]{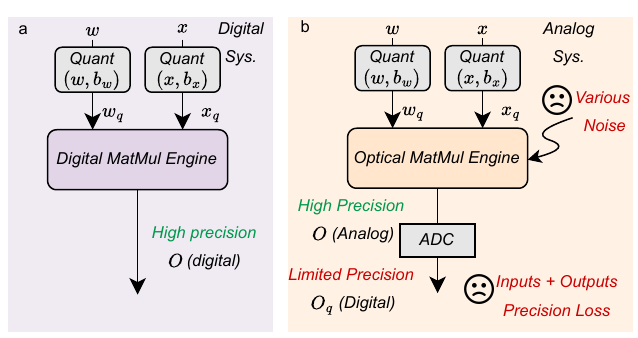}
    \vspace{-5pt}
    \caption{Comparison between (a) the low-precision digital matrix multiplication (MatMul) engine and (b) the low-precision optical MatMul engine. The optical MatMul engine suffer from various non-idealities during computation and hold precision loss for both inputs and outputs.
    }
    \label{fig:compare_prec}
    \vspace{-10pt}
\end{figure}

Ensuring functional correctness is a fundamental requirement of computing hardware.
However, analog photonic computing inherently faces robustness challenges due to two primary factors, as depicted in Fig.~\ref{fig:compare_prec}: 
(1) Various non-ideal conditions include process variations, device noises, environmental factors, and limited endurance.
(2) Limited precision of inputs and outputs arises from the finite control resolution and the substantial overhead associated with high-precision DACs and ADCs.
Unlike low-precision electronics digital computing, as shown in Fig.~\ref{fig:compare_prec}.a, photonic computing uniquely suffers from output precision loss in the conversion of analog outputs back to the digital domain, where the ADC precision is typically unsatisfactory compared to the output precision.
To understand the mismatch between an ADC and output precision, consider a straightforward scenario in which a photonic computing engine executes a dot product operation of length $N$.
Note that here we consider an ideal case where the precision constraints are on the weights and activations, and the photonic computing engine can be treated as a dot-product engine based on a scalar product.\footnote{The real case can be more intricate. For example, low precision may be reflected in device control variables, and photonic computing cannot be easily treated based on scalar product.} 
The multiplication of a $B_w$-bit signed weight by a $B_x$-bit signed activation results in a product with $B_w + B_x -2 + \log_2 N$ magnitude bits and one sign bit. 
Here, the additional $log_2 N$ bits of precision stem from the addition of $N$ scalar products. 
However, achieving this desired precision often exceeds the capabilities of affordable ADCs, typically within 8 bits~\cite{nahmias2019photonic}.
For instance, a 4-bit weight and 4-bit activation in a length-$32$ dot product computation would ideally yield an output precision of 13 bits.
The 5-bit overhead underscores a notable difference between photonic computing and traditional low-precision arithmetic units like int8 GPU tensor cores where outputs are preserved in high precision~\cite{choquette2021nvidia}, raising a unique precision issue.

\subsubsection{Co-design Progress}
Here, we explore efforts to mitigate the precision issue and the impact of noise on accuracy.
Regarding precision issues, previous efforts have often drawn inspiration from quantization, either employing Quanztiation-aware Training (QAT)~\cite{gu2020roq,zhu2022elight,kirtas2023mixed,wang2022integrated, zhu2023multi}, or Post-training Quantization (PTQ)~\cite{zhang2021training}. 
These approaches enable neural networks to inherently tolerate low-precision arithmetic, thereby mitigating the impact of precision limitations.
Specifically, Gu et al. first developed the QAT flow by directly optimizing low-precision device control signals in the discretized space~\cite{gu2020roq}.
However, the aforementioned quantization techniques are insufficient to compensate for the computation information loss due to the low-precision ADC. This is because errors occur in each partial result, and they finally accumulate when we tile a large matrix workload.
Some work in analog computing specifically tune the ADC reference voltage to balance the tradeoff between dynamic quantization range and solution~\cite{rekhi2019analog}. Furthermore, weight and input slicing are proposed as strategies to manage the trade-off in ADC resolution~\cite{andrulis2023raella}.

For noise mitigation, noise-aware training has emerged as a widely used technique to enhance resilience~\cite{zhao2019design,gu2020roq, sludds2022delocalized}. This approach involves introducing noises and variations during the training process to enhance the noise resilience of ONNs.
Previous studies have focused on modeling various noise sources~\cite{fang2019design,sludds2022delocalized}, such as dynamic noise, static manufacturing variation, and thermal crosstalk, then incorporated them into the training process.
Furthermore, explicit robustness optimization terms can be integrated into the training process to further enhance robustness against noise.
For instance, Ref~\cite{gu2020roq} estimates the noise sensitivity of weights and applies protective regularization terms to sensitive weights during optimization.
Similarly, Zhu et al. introduced an additional regularization term on the phase magnitude in an MZI mesh to reduce the crosstalk concerns~\cite{zhu2020countering}.
Additionally, knowledge distillation strategies, as employed in Ref~\cite{gu2021o2nn}, can be employed to guide the optimization of noisy student ONN models under the guidance of a noise-free teacher model. This approach significantly enhances the robustness of models against static process variation and dynamic input signal noises.

However, noise patterns can be highly intricate, rendering the analysis and modeling of these patterns both challenging and time-consuming.
In response to this challenge, some studies have explored training ONNs directly with noisy physical responses from real chips, eliminating the need for explicit noise modeling~\cite{feng2022compact,wright2022deep,zhan2024physics}.
This approach, known as physical neural network training, has been introduced in detail in Section~\ref{subsec:onn_training}.
Moreover, on-chip training is another way to recover accuracy by inherently modeling on-chip noise during training, which is further explored in Section~\ref{subsec:learnability}.

Besides exploring the inherent noise tolerance of neural networks, another straightforward direction is to compress the on-chip noise levels. 
Specifically, some work advocate employing device-level and circuit-level design space exploration to mitigate the impact of process variation and crosstalk~\cite{sunny2021crosslight, mirza2021silicon}. 
Ref~\cite{zhu2022elight} suggests avoiding frequent reprogramming on PCM cells by enhancing the similarity of mapping weights to mitigate early wear-out. 
Additionally, reducing the number of active devices in PICs can also decrease on-chip noise, as noise-induced errors typically correlate positively with the number of noise sources.
Therefore, pruning redundant devices~\cite{gu2022adept}, or weight blocks~\cite{gu2020towards, feng2022compact}, brings significant noise robustness improvement.
\subsection{Self-learnability}
\label{subsec:learnability}
\subsubsection{Issue}
The adaptability or trainability of photonic analog computing platforms is another major challenge in their practical application.
The significance of self-learnability arises from three main aspects, as shown in Fig.~\ref{fig:learnable}.
Firstly, on-chip training enhances the adaptability of optical hardware when working conditions drift, or when workloads change.
Secondly, the realization of on-chip training can unlock many important edge learning applications such as local online learning, transfer learning, and lifelong learning.
Moreover, as discussed in Section~\ref{subsec:onn_training}, on-chip training can address robustness issues \emph{in-situ} and bridge the gap between simulation and real implementation.

\begin{table*}[ht]
\centering
\caption{
Comparison of different on-chip training protocols on gradient estimation methods and notable hardware requirements. We also show the ONN scale they can handle in terms of low ($\sim100$), medium ($\sim1000$), and high ($>10$k).
}
\label{tab:compare_training}
\resizebox{0.95\textwidth}{!}{
\begin{tabular}{c|c|c|c}
\hline
On-chip training protocol & Backward pass                                  & Hardware requirements & ONN scale \\
\hline
Gradient-free Opt.~\cite{zhou2019chip,shen2017deep,zhang2021efficient} & No gradient & Precise device control & Low \\
Direct feedback alignment~\cite{filipovich2022silicon} & Estimated gradient with random projection & Random projection unit & Low \\
Adjoint-variable method~\cite{hughes2018training, pai2023experimentally} & True gradient with adjoint method & Bidirectional I/O, per device monitor & Low \\
Zeroth-order Opt.~\cite{bandyopadhyay2022single,gu2020flops, gu2021efficient} & Estimated gradient with finite difference method & Forward-only & Medium \\
First-order Opt.~\cite{gu2021l2ight} & True first-order gradient & Bidirectional I/O & High \\
\hline
\end{tabular}
}
\vspace{-10pt}
\end{table*}

\subsubsection{Co-design Progress}
Achieving self-learnability on neuromorphic photonic processors is indeed challenging due to various on-chip restrictions, such as inaccessible gradients for control variables and a lack of full observability for in-situ light fields. 
Consequently, there is a demand for developing hardware-friendly learning algorithms that can operate within these constraints while maintaining feasibility.

We summarize recent on-chip training protocols, either validated through simulation or real demonstration, in Table~\ref{tab:compare_training}.
Generic gradient-free optimization methods, e.g., evolutionary algorithms and brute-force device tuning~\cite{zhou2019chip,shen2017deep,zhang2021efficient}, have been used to optimize on-chip parameters with no gradient information involved.
Nevertheless, these approaches encounter difficulties when scaling to large-scale models with slow convergence and poor stability.
Besides the gradient-free approach, investigation into the direct feedback alignment (DFA) training algorithm is further explored in Ref~\cite{filipovich2022silicon} for \emph{in-situ} training~\cite{nokland2016direct, launay2020direct}. 
DFA propagates errors through fixed random feedback projections from the output layer to each hidden layer in parallel, thereby obviating the necessity for the sequential backpropagation of gradients.
Another approach, the adjoint variable method, was introduced to perform on-chip backpropagation by computing the gradients \emph{in-situ} with per-device optical field monitors~\cite{hughes2018training}, and recent work has experimentally demonstrated the concept~\cite{pai2023experimentally}.
However, scalability remains a significant concern due to considerable hardware support overhead, such as the need for per-device monitoring.

\begin{figure}[htb]
    \centering
    \vspace{-5pt}
\includegraphics[width=0.95\columnwidth]{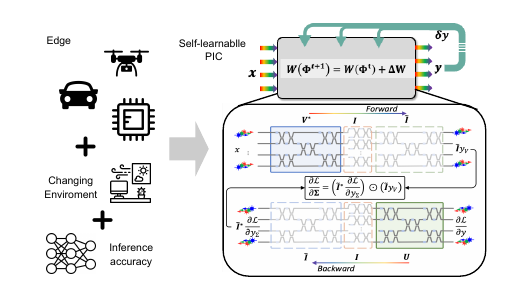}
    \vspace{-3pt}
    \caption{The self-learnable ONN paradigm exemplified by a single scalable ONN on-chip training framework with \emph{in-situ} gradient calculation~\cite{gu2021l2ight}.
    }
    \label{fig:learnable}
    \vspace{-10pt}
\end{figure}

On-chip training protocols based on forward-only zeroth-order optimization aim to approximate gradients using the finite difference method, perturbing parameters with small random values for gradient estimation. 
Instead of adjusting individual on-chip parameters separately, Bandyopadhyay et al. proposed a method that samples perturbations for all parameters and shares them among all training instances in a single iteration~\cite{bandyopadhyay2022single}. Similarly, FLOPS, proposed in Ref~\cite{gu2020flops}, shares the sampled perturbations among single mini-batches instead of all mini-batches in the same interaction, showing better convergence than Ref~\cite{bandyopadhyay2022single}.
To further enhance the scalability of zeroth-order training protocols, MixedTrain proposed in Ref~\cite{gu2021efficient} partitions PICs into passive and active regions and trains only a small subset of active devices in each iteration, ensuring the capacity to handle a larger scale ONN compared to FLOPS.
The zeroth-order-based approach shows the training scale to thousands of MZIs as shown in Ref~\cite{gu2020flops, gu2021efficient}, while it is still not enough when dealing with modern-size machine learning models.
L$^2$ight introduces a subspace optimization algorithm and develops a method for \textit{in-situ} calculation of first-order subspace gradients~\cite{gu2021l2ight}, as shown in Fig.~\ref{fig:learnable}. Additionally, a multi-level hardware-aware sparse training method is employed to boost training efficiency. 
This study demonstrates the first instance of training a million-parameter-level ONN, showcasing exceptional stability.
\section{Outlook and Conclusion}\label{section6}
\label{sec:outlook}

In this review, we have outlined recent advancements in photonic-electronic integrated circuits for computing and AI tasks, covering the progress made at various levels, including the device, circuit, architecture, and system level, as well as progress in cross-layer software-hardware co-design strategies.
However, more technical challenges need to be addressed in the future to enable the practical application of photonic computing.
Here, we point to possible research directions.

\subsubsection{Device/new material innovation} 
As highlighted in this review, the characteristics of optical components directly impact the performance of photonic-electronic computing systems.
Recent research in emerging materials and novel devices provide new opportunities for enhancing the performance of PICs from the device level. For instance, heterogeneous modulators, leveraging technologies such as III-V MOSCAP and 2-D materials~\cite{hiraki2017heterogeneously,sorianello2018graphene}, demonstrate high modulation efficiency, gigahertz-level cut-off frequencies, and sub-picojoule-per-bit power consumption, allowing a compact layout with high energy efficiency.
Additionally, dataflow management in PIC-based computing, which includes both digital and analog systems, primarily depends on electrical memory due to the lack of equivalent components in the photonic domain. Previous studies have demonstrated that non-volatile materials and devices have the capability to store data, including pre-trained weight parameters~\cite{feldmann2019all,feldmann2021parallel}.
Significantly, non-volatile photonic memory, in contrast to conventional electronic memory, serves not only as a data storage solution but also functions as a computational unit for `in-memory' computing.
Nevertheless, due to the limited switching frequency of current non-volatile materials, such as PCMs, scenarios requiring dynamic data reading and writing (particularly in high-performance digital processing units) demand high-speed, reconfigurable materials for optical memory.

Beyond the innovation in optical components for computing reviewed in Section~\ref{subsubsec:efficent_ptc}, optimizing other aspects of PICs also presents significant interest. Potential research directions encompass, but are not limited to (1) on-chip light sources to eliminate the complex fiber packaging and alignment processes; (2) integrated frequency combs as WDM sources~\cite{feldmann2021parallel}; (3) customized nonlinear units for on-chip implementation of programmable activation functions; and (4) efficient ADC/DACs and photodetectors, specifically designed for optical computing architectures~\cite{meng2021electronic}.

\subsubsection{Advanced photonics and electronics integration}

PICs for computing require on-chip interfacing with diverse electronic components, including signal conversion circuits and control systems, leading to substantial data communication between the electronic and photonic parts. Currently, wirebonding is a widely adopted strategy for E-O interconnects in academia. This method is particularly convenient and cost-effective for small-scale networks with few active devices, as it requires only the routing of driving signals to the edges of photonic chips, followed by connections to peripheral electronic units via wire bonds. However, as PICs scale up, managing metal wire routing within constrained chip areas presents a challenge. The beachfront bottleneck of photonic chips and electrical control units restricts the number of connections that can be implemented due to the limited chip perimeter. Furthermore, wirebonding can constrain the bandwidth of high-frequency devices owing to impedance mismatch.

To address this interconnect challenge, advanced integration methods can be employed, aiming to reduce interconnect complexity and improve overall system performance \cite{Lightmatter2024}. Flip-chip bonding directly connects two dies by soldering or using conductive bumps to align matching electrical pads on their surfaces. This approach allows flexible floorplanning and pad placement beyond chip edges, enabling higher interconnect density and reducing parasitic impedance. Given the fabrication process and chip sizes, typical integration strategies involve flip-chip bonding single or multiple heterogeneous dies (such as ADC/DAC, memory, microcontroller, FPGA, ASIC, etc.) onto a photonics chip which also serves as an interposer. However, this approach will invariably introduce large temperature excursions on the PIC varying with the workload. To address this issue, one strategy involves developing temperature-resilient photonic platforms \cite{sahni2024beyond}. Additionally, the heat spreader and through-silicon via (TSV) technique enables more flexible thermal management within chips \cite{tang2015thermal,refai2020lidless,shastri2021photonics, bogaerts2018silicon}.

Monolithic fabrication processes that integrate both photonic and electronic components on a single substrate also have shown promising results~\cite{stojanovic2018monolithic, giewont2019300}. While this technology enables higher integration levels, bandwidth, and energy efficiency, significant potential remains for the optimization of fabrication processes, improving yield, and the development of comprehensive PDKs. Besides, further breakthroughs, such as integration with more advanced CMOS nodes and 2.5-D/3-D electro-optical integration, are anticipated to enhance the performance of photonic chips.

\subsubsection{Scale to large models and advanced tasks}
Photonic computing still faces challenges concerning scalability when moving to support large models.
Several directions can be explored.
First, the continued exploration of domain-specific PTCs which trade universality for higher scalability, as indicated by previous research efforts~\cite{gu2020towards, feng2022compact, zhu2022space, xiao2021large, gu2023m3icro, li2023photofourier}.
Second, there exists a substantial potential to further improve computational density by engineering tailored, compact photonic devices, such as multi-operand modulators and metasurface-based devices. 
Third, the exploitation of the unique characteristics of photonics, such as wavelength, time, or mode-division multiplexing, can enable the reuse of hardware for a higher degree of parallel operations. 
Lastly, the development of a deep understanding of the workload of evolving machine learning models is crucial for designing suitable PICs.
For instance, attention-based Transformer models introduce dynamic matrix multiplication, challenging previous PTC designs optimized for CNNs.
Recent advancements propose dynamically-operated PTC design to handle dynamic matrix multiplication efficiently, ensuring optimal performance across diverse AI applications.

\subsubsection{On-chip ONN training protocol}
The mainstay of ONN training predominantly relies on simulation, while existing on-chip training experimental demonstrations remain at a small scale with notable overheads. 
A stable and efficient on-chip training scheme is highly coveted, necessitating breakthroughs in both hardware and training algorithms. These breakthroughs may include innovations such as light-field-driven nonvolatile materials and advancements in training algorithms.

\subsubsection{Cross-layer co-design and electronic-photonic design automation (EPDA)}

Cross-layer efforts open up additional opportunities to optimize the performance of ONNs, as discussed in Section~\ref{section5}. 
With the growing complexity of photonic-electronic hardware platforms, exploring EPDA becomes critical for enhancing productivity and efficiency, such as exploring automatic circuit layout generation and fast photonic circuit simulation~\cite{gu2022neurolight}. 

\subsubsection{System-level simulator}
As a multidisciplinary emerging area, it is crucial to have a comprehensive simulator framework for evaluating the performance of optical computing systems.
Ideal simulators should support seamless integration of new optical hardware, offer automatic algorithms for hardware mapping, and provide evaluation of chip-level performance.
Such tools will facilitate fair and straightforward evaluation across different optical circuit designs, helping identify system bottlenecks and guiding further optimizations.

\subsubsection{Optical AI software stack}
Developing specialized compilers and instruction sets tailored for photonic computing architectures is necessary to enable the integration of novel optical hardware into the mainstream AI software stack, as envisioned in Ref~\cite{ferreira2020primer}. While existing deep learning frameworks such as PyTorch and TensorFlow can still serve as the front end, the compiler component must be adapted to generate machine code optimized for the unique photonic accelerators. This ensures efficient utilization of the optical hardware. Moreover, we anticipate open-source efforts in this direction to facilitate the progress of the optical AI software stack.

In conclusion, photonic-electronic integrated circuits stand out for their exceptional advantages in computing—in terms of low latency, high bandwidth, and energy efficiency—and exhibit a high potential to overcome the insurmountable bottlenecks of electronic computing. 
Nevertheless, whether for digital or analog computing, the PIC scale of reported work remains limited, and further improvements in scalability are essential. Continuous research should aim to enhance throughput via innovative devices, architectural improvements, specialized training algorithms, and hardware-software co-design strategies. Additionally, efforts should focus on reducing power consumption and the costs associated with the E-O interface, in order to rival the performance of state-of-the-art electronic computing systems. This ambition, however, does not imply that the objective of optical computing is to surpass digital-electronic processors in all metrics and replace them. From an industry perspective, optical computing needs to identify niches where it excels over its electronic counterparts. A promising short-term application involves leveraging the high parallelism of optical computing to develop MVM processing units and accelerators for neural networks, an area that has been extensively studied. If the PIC scale can be well aligned with its tasks (including considerations for multiplexing), and use efficient E-O interfaces, then operation at a rate of one MVM per nanosecond could be achieved and offer significant advantages in high-throughput applications.
With sustained innovation and effort, integrated photonics is poised to become a pivotal emerging technology, satisfying the escalating societal demand for high-performance computing and hardware acceleration in AI applications over the long term.


\bibliographystyle{IEEEtran}

\bibliography{refer} 



 




\vfill

\end{document}